\documentclass[a4paper,12pt]{article}
\pdfoutput=1
\usepackage{graphicx,rotating,amssymb,amsmath,slashed,enumerate}
\usepackage[sort&compress,numbers,merge]{natbib}
\ifx\pdfoutput\undefined
\usepackage[dvips,bookmarks]{hyperref}	
\else
\usepackage{hyperref}	
\fi
\hypersetup{colorlinks,bookmarksopen,bookmarksnumbered,citecolor=verdes,
linkcolor=verdec,pdfstartview=FitH,urlcolor=rossos}

\def\hhref#1{\href{http://arxiv.org/abs/#1}{#1}} 

\usepackage{multicol}
\usepackage{color}
\definecolor{rosso}{cmyk}{0,1,1,0.4}
\definecolor{rossos}{cmyk}{0,1,1,0.55}
\definecolor{rossoc}{cmyk}{0,1,1,0.2}
\definecolor{blu}{cmyk}{1,1,0,0.3}
\definecolor{blus}{cmyk}{1,1,0,0.6}
\definecolor{bluc}{cmyk}{1,1,0,0.1}
\definecolor{verde}{cmyk}{0.92,0,0.59,0.25}
\definecolor{verdec}{cmyk}{0.92,0,0.59,0.15}
\definecolor{verdes}{cmyk}{0.92,0,0.59,0.4}

\font\tenrsfs=rsfs10 at 12pt
\font\sevenrsfs=rsfs7
\font\fiversfs=rsfs5
\newfam\rsfsfam
\textfont\rsfsfam=\tenrsfs
\scriptfont\rsfsfam=\sevenrsfs
\scriptscriptfont\rsfsfam=\fiversfs
\def\mathscr#1{{\fam\rsfsfam\relax#1}}

\newcommand{\TeV}{\,{\rm TeV}}
\newcommand{\eV}{\,{\rm eV}}

\def\circa#1{\,\raise.3ex\hbox{$#1$\kern-.75em\lower1ex\hbox{$\sim$}}\,}

\def\LUX{{\sc Lux}} 
\def\CRESST{{\sc Cresst-II}}

\newcommand{\beq}{\begin{equation}}
\newcommand{\eeq}{\end{equation}}

\def\circa#1{\,\raise.3ex\hbox{$#1$\kern-.75em\lower1ex\hbox{$\sim$}}\,}
\makeatletter

\def\art{\@ifnextchar[{\eart}{\oart}}
\def\eart[#1]#2#3#4#5#6{{\rm #2}, {#3 #4} {\rm (#6) #5} [{\hhref{#1}}]}
\def\hepart[#1]#2{{\rm #2, \hhref{#1}}}
\newcommand{\oart}[5]{{\rm #1}, {#2 #3} {\rm (#5) #4}}

\newcounter{alphaequation}[equation]
\def\thealphaequation{\theequation\hbox to
0.6em{\hfil\alph{alphaequation}\hfil}}
\def\eqnsystem#1{
\def\@eqnnum{{\rm (\thealphaequation)}}
\def\@@eqncr{\let\@tempa\relax \ifcase\@eqcnt \def\@tempa{& & &} \or
  \def\@tempa{& &}\or \def\@tempa{&}\fi\@tempa
  \if@eqnsw\@eqnnum\refstepcounter{alphaequation}\fi
\global\@eqnswtrue\global\@eqcnt=0\cr}
\refstepcounter{equation} \let\@currentlabel\theequation \def\@tempb{#1}
\ifx\@tempb\empty\else\label{#1}\fi
\refstepcounter{alphaequation}
\let\@currentlabel\thealphaequation
\global\@eqnswtrue\global\@eqcnt=0 \tabskip\@centering\let\\=\@eqncr
$$\halign to \displaywidth\bgroup \@eqnsel\hskip\@centering
$\displaystyle\tabskip\z@{##}$&\global\@eqcnt\@ne
\hskip2\arraycolsep\hfil${##}$\hfil& \global\@eqcnt\tw@\hskip2\arraycolsep
$\displaystyle\tabskip\z@{##}$\hfil
\tabskip\@centering&\llap{##}\tabskip\z@\cr}
\def\endeqnsystem{\@@eqncr\egroup$$\global\@ignoretrue} \makeatother

\usepackage{relsize}
\newcommand{\alphaD}		{\alpha_{\mathsmaller{D}}}

\newcommand{\fVD}		{f_{\mathsmaller{\VD}}}
\newcommand{\gD}			{g_{\mathsmaller{D}}}
\newcommand{\gDtilde}	{\tilde{g}_{\mathsmaller{D}}}
\newcommand{\gSM}		{g_{\mathsmaller{\rm SM}}}
\newcommand{\gSMtilde}	{\tilde{g}_{\mathsmaller{\rm SM}}}
\newcommand{\gVD}		{g_{\mathsmaller{\VD}}}
\newcommand{\MDM}		{M_{\mathsmaller{\rm DM}}}
\newcommand{\mVD}		{m_{\mathsmaller{\VD}}}
\newcommand{\nVD}		{n_{\mathsmaller{\VD}}}
\newcommand{\sSM}		{s_{\mathsmaller{\rm SM}}}
\newcommand{\tauBBN}		{\tau_{\mathsmaller{\rm BBN}}}
\newcommand{\tauVD}		{\tau_{\mathsmaller{\VD}}}
\newcommand{\TD}			{T_{\mathsmaller{D}}}
\newcommand{\TX}			{T_{\mathsmaller{X}}}
\newcommand{\TSM}		{T_{\mathsmaller{\rm SM}}}
\newcommand{\xD}			{x_{\mathsmaller{D}}}
\newcommand{\VD}			{{V_{\mathsmaller{D}}}}
\newcommand{\pss}		{{s_{ps}}}
\usepackage{cleveref} 
\usepackage[titles]{tocloft}

\begin{document}
\begin{center}
{\footnotesize
CERN-TH-2016-255 \hfill IFT-UAM/CSIC-16-144 \hfill NIKHEF-2016-062}
\end{center}
\color{black}

\begin{center}
\textbf{\Huge
Dark Matter's secret \textit{liaisons:}\\[4mm]} 
{\Large\bf  phenomenology of a dark $\boldsymbol{U(1)}$ sector with bound states
}

\medskip
\bigskip\color{black}\vspace{0.6cm}

{
{\large\bf Marco Cirelli}\ $^a$,
{\large\bf Paolo Panci}\ $^{b,c}$,
{\large\bf Kalliopi Petraki}\ $^{a,d}$,\\[3mm]
{\large\bf Filippo Sala}\ $^a$,
{\large\bf Marco Taoso}\ $^e$
}
\\[7mm]
{\it $^a$ \href{http://www.lpthe.jussieu.fr/spip/index.php}{Laboratoire de Physique Th\'eorique et Hautes Energies (LPTHE)},\\ UMR 7589 CNRS \& UPMC,\\ 4 Place Jussieu, F-75252, Paris, France}\\[3mm]
{\it $^b$ \href{http://th-dep.web.cern.ch/welcome-page}{CERN Theoretical Physics Department}, CERN, \\ 
Case C01600, CH-1211 Gen\`eve, Switzerland}\\[3mm]	
{\it $^c$ \href{http://www.iap.fr}{Institut d'Astrophysique de Paris}, \\ UMR 7095 CNRS, Universit\'e Pierre et Marie Curie, \\
98 bis Boulevard Arago, Paris 75014, France}\\[3mm]
{\it $^d$ \href{https://www.nikhef.nl/en/}{Nikhef}, Science Park 105, 1098 XG Amsterdam, The Netherlands}\\[3mm]
{\it $^e$ \href{https://www.ift.uam-csic.es/en}{Instituto de F\'isica Te\'orica (IFT)} UAM/CSIC,\\ calle Nicol\'as Cabrera 13-15, 28049 Cantoblanco, Madrid, Spain}
\end{center}

\bigskip

\centerline{\large\bf Abstract}
\begin{quote}
\color{black}\large

Dark matter (DM) charged under a dark $U(1)$ force appears in many extensions of the Standard Model, and has been invoked to explain anomalies in cosmic-ray data, as well as a self-interacting DM candidate. In this paper, we perform a comprehensive phenomenological analysis of such a model, assuming that the DM abundance arises from the thermal freeze-out of the dark interactions. We include, for the first time, bound-state effects both in the DM production and in the indirect detection signals, and quantify their importance for {\sc Fermi}, {\sc Ams-02}, and CMB experiments. We find that DM in the mass range 1 GeV to 100 TeV, annihilating into dark photons of MeV to GeV mass, is in conflict with observations. Instead, DM annihilation into heavier dark photons is viable. We point out that the late decays of multi-GeV dark photons can produce significant entropy and thus dilute the DM density. This can lower considerably the dark coupling needed to obtain the DM abundance, and in turn relax the existing constraints.

\end{quote}

\clearpage

\setlength{\cftbeforesecskip}{1.3ex}
\noindent\makebox[\linewidth]{\rule{\textwidth}{1pt}} 

\vspace{-3ex}
\tableofcontents
\noindent\makebox[\linewidth]{\rule{\textwidth}{1pt}}

\section{Introduction}
\label{sec:introduction}

The physics of the {\em dark sector}, comprising most notably the Dark Matter (DM) particle responsible for incontrovertible astrophysical and cosmological evidence, is still largely, and aptly, obscure (see e.g.~\cite{Bertone:2004pz,Feng:2010gw} for reviews).
Not only the mass of the DM particle is undetermined, with viable possibilities ranging from tiny fractions of one eV to hundreds of TeV, but perhaps most importantly, its interaction properties are so far unknown (beyond, of course, the coupling with gravity which is at the origin of the evidence mentioned above). Historically, much of the attention has concentrated on WIMP scenarios, in which the DM particle -- often assumed, for simplicity, to consist of just one species -- is charged under the well-known weak interactions of the Standard Model (SM). Under this logic, the matter content of the theory is enlarged but the gauge force sector is left unchanged. The relative predictivity of these scenarios, and their appealing embedding within broader theory constructions such as supersymmetry, have made them paradigmatic for the field in the past decades.

On the other hand, it is interesting to entertain the possibility that the DM may be charged under a  new {\em dark force}, carried by a new {\em dark mediator}. We will consider, specifically, DM coupled to a dark Abelian gauge interaction $U(1)_D$, and allow for the dark gauge boson, which we shall call the dark photon, to be massive.  Such a force couples DM with itself, making it potentially collisional. It also generically allows DM to couple to the Standard Model (SM), via the renormalisable kinetic mixing of the $U(1)_D$ with the hypercharge~\cite{Holdom:1985ag,Foot:1991kb}; this gives rise to direct and indirect detection signals, and has important implications for cosmology.

Models of this sort have been investigated in many occasions in the past, based both on theoretical~\cite{Kors:2004dx, Feldman:2006wd, Fayet:2007ua, Goodsell:2009xc, Morrissey:2009ur, Andreas:2011in, Goodsell:2011wn, Fayet:2016nyc}  and phenomenological motivations. The latter includes, for instance, the long-standing possibility that sizeable DM self-interactions can solve some worrying inconsistencies between the predictions of collisionless cold DM and the observed large-scale structure of the Universe~\cite{Spergel:1999mh}, that appear, for example, in the sizes and numbers of the galactic satellite haloes, and in the DM density profiles in the center of galaxies~\cite{Feng:2008mu, Loeb:2010gj, Rocha:2012jg, Peter:2012jh, Zavala:2012us, Vogelsberger:2014pda, Schneider:2016ayw}. 
The same class of models was also invoked in the wake of the discovery of a positron excess in cosmic rays by the {\sc Pamela} satellite~\cite{Pospelov:2007mp, Cholis:2008vb, Pospelov:2008jd, Nelson:2008hj, Cholis:2008qq, Feldman:2008xs, Bergstrom:2008ag, ArkaniHamed:2008qn}; the specific implementations featured a DM particle that annihilates into mediators with a relatively small mass, below the threshold for $p \bar p$ production. This feature allowed to explain the leptophilic nature of the annihilation and, at the same time, to provide the very sizeable enhancement of the annihilation cross-section -- the so-called Sommerfeld enhancement~\cite{Hisano:2003ec, Hisano:2004ds, Belotsky:2004st, Hisano:2006nn, Cirelli:2007xd, ArkaniHamed:2008qn, Feng:2009hw, Feng:2010zp} -- that was needed to fit the data. 
Later on, models in this class were constructed to explain the {\sc Fermi} Galactic Center GeV excess~\cite{Goodenough:2009gk,Vitale:2009hr, Hooper:2010mq, Hooper:2011ti,Abazajian:2012pn,Daylan:2014rsa,TheFermi-LAT:2015kwa}. In these constructions, the existence of dark mediators allows to obtain smoother gamma-ray spectra and to avoid other constraints~\cite{Hooper:2012cw, Martin:2014sxa,Boehm:2014bia, Abdullah:2014lla, Berlin:2014pya, Cline:2014dwa, Liu:2014cma, Cline:2015qha, Elor:2015tva}.

As seen from the above, the suitability of the $U(1)_D$ model to address certain problems 
relies often on the smallness of the force mediator mass. More recently, it has been realised in the context of DM phenomenology that long-range interactions -- interactions mediated by light or massless force carriers -- imply the existence and formation of DM \emph{bound states}. 
This applies both to dark force scenarios, as well as to WIMPs, provided that the force carriers (the dark mediator or the SM Weak bosons) are much lighter than the DM particles. After the pioneering work of refs.~\cite{Pospelov:2008jd, MarchRussell:2008tu,Shepherd:2009sa}, the issue has been investigated more systematically in a series of recent studies~\cite{vonHarling:2014kha, Petraki:2015hla, Pearce:2013ola, Petraki:2014uza, Pearce:2015zca, Laha:2013gva, Laha:2015yoa, An:2015pva, An:2016gad, An:2016kie, Kouvaris:2016ltf, Bi:2016gca, Kim:2016kxt, Kim:2016zyy, Asadi:2016ybp, Petraki:2016cnz, Johnson:2016sjs, Liew:2016hqo}. 
Dark matter with no particle-antiparticle asymmetry, coupled to long-range interactions can form unstable particle-antiparticle bound states. 
Their formation and decay has an important impact on DM phenomenology: it provides an additional annihilation channel, which affects the relic abundance~\cite{Petraki:2014uza}, as well as the radiative signals looked for in indirect detection strategies~\cite{An:2016gad, An:2016kie, Kouvaris:2016ltf}. The modifications can be significant, and need to be taken into account. 
 
\medskip

The goal of this paper is to perform, for the first time, a comprehensive and self-consistent analysis of the $U(1)_D$ model that will include the impact of bound states. We compute precisely the DM relic density, incorporating the effect of bound-state formation (BSF) and decay, which reduces the predicted DM coupling~\cite{vonHarling:2014kha}. We analyze the constraints on the parameter space of the model that come from beam dump experiments, supernova cooling, DM direct detection and Big Bang Nucleosynthesis. Then we consider the most promising DM indirect detection probes: the gamma-ray signals from the Milky Way galactic halo and from dwarf galaxies, the antiproton measurements by the {\sc Ams-02} experiment and the anisotropies of the Cosmic Microwave Background (CMB).  Recently, two related studies with some overlap with ours 
have appeared in the literature~\cite{An:2016gad,Bringmann:2016din}; our results broadly agree with theirs, but our analysis has a wider scope, uses some different ingredients and points out new effects.\footnote{
The main differences with~\cite{An:2016gad} are the following. 
i) We take into account the effect of BSF on the DM relic density, hence on the DM coupling to the dark photon.
ii) While they essentially focus their indirect detection analysis on $\gamma$-rays from the Galactic Center (GC), we employ a wider array of probes. 
iii) We avoid relying on the GC as it is a challenging region (where actually an excess is currently discussed) and therefore it is less suitable for deriving constraints. 
iv) We include the Inverse Compton Scattering component in the computation of the $\gamma$-ray spectra: this component significantly increases the flux at small energies, say $10^{-2}$ times the DM mass, and it cannot therefore be neglected especially if one wants to derive constraints on DM as heavy as 10 to 100 TeV using the {\sc Fermi} data that extend to hundreds of GeVs at most. 
We will come back on these points in detail in the following sections.}${}^,$\footnote{
The very recent study in~\cite{Bringmann:2016din} differs from ours in that it does not consider our main topic, i.e.~the formation of bound states, while it focuses instead on DM self-interactions. Moreover, we differ in some of the probes that we consider to derive limits: we do not rely on {\sc Ams-02} positron measurements (because they are based on an unknown astrophysical background, at the origin of the positron rise), while they do not consider the {\sc Ams-02} antiprotons and the {\sc Fermi} galactic halo measurements. Finally, the parameter space of their analysis is somewhat different from ours. We consider a larger range of dark photon masses (while they focus on light mediators), and point out important implications from the decay of the cosmological abundance of heavier dark photons.} 
For more work related to the indirect detection of this kind of models, see also~\cite{Bell:2016fqf,Kouvaris:2016ltf,Feng:2015hja,Feng:2016ijc}.

\medskip

The rest of this paper is organized as follows. 
In~\cref{sec:stage}, we specify the details of the particle physics model, and we recall the main features associated with the formation of bound states. We also set our notation, and define the parameter space on which we focus. 
In~\cref{sec:cosmo}, we expand on the cosmology of the model. We detail the computation of the DM relic abundance, describe the cosmological evolution of the DM and dark radiation after freeze-out, consider constraints arising from BBN, and quantify how late decays of dark photons might dilute the DM abundance. 
In~\cref{sec:DMID}, we present the constraints we obtain from indirect detection, based on gamma-ray measurements by {\sc Fermi} and anti-proton searches by {\sc Ams-02}, and the constraints we obtain from CMB, based on measurements by {\sc Planck}.
Finally, in~\cref{sec:conclusions}, we summarize and we present our conclusions.

\section{Setting the stage}
\label{sec:stage}

We consider DM in the form of Dirac fermions $X$ coupled to a dark force $U(1)_D$ carried by a dark photon $\VD$, that mixes kinetically with the hypercharge $U(1)_Y$. The Lagrangian of the dark sector reads
\beq
{\cal L} = \bar{X}(i \slashed{D} - \MDM)X 
- \frac{1}{4}{F_{\mathsmaller{D}}}_{\mu\nu}F_{\mathsmaller{D}}^{\mu\nu}
- \frac{1}{2} \, \mVD^2 \VD_{\mu} \VD^{\mu} 
- \frac{\epsilon}{2c_w}{F_{\mathsmaller{D}}}_{\mu\nu}F_{\mathsmaller{Y}}^{\mu\nu} \,,
\label{eq:L}
\eeq
where the covariant derivative for $X$ is $D^\mu = \partial^\mu + i g_d \VD^\mu$, and $F_{\mathsmaller{D}}^{\mu\nu} = \partial^\mu \VD^\nu - \partial^\nu \VD^\mu$. The mass $\mVD$ of the dark photon field may arise either via the St\"uckelberg mechanism or a dark Higgs mechanism. In~\cref{sec:cosmo}, we shall indicate the differences in the cosmological evolution implied by the two mechanisms, whenever they arise.

Hence the parameters of the system are: $\MDM$ the DM particle mass, $\mVD$ the dark photon mass, $\alphaD = g_d^2/4\pi$ the dark fine structure constant and $\epsilon$ the kinetic mixing parameter. We will restrict to the case in which $\MDM > \mVD$, so that DM annihilations into two dark photons are always kinematically possible. As described in~\cref{sec:relic}, we will determine $\alphaD$, with respect to $\MDM$ and $\mVD$, by requiring that the observed DM density arises from the thermal freeze-out of the $X$ fermions in the hidden sector.

\begin{figure}[t]
\centering
\begin{minipage}{0.31 \textwidth}
\includegraphics[width= 1.25 \textwidth]{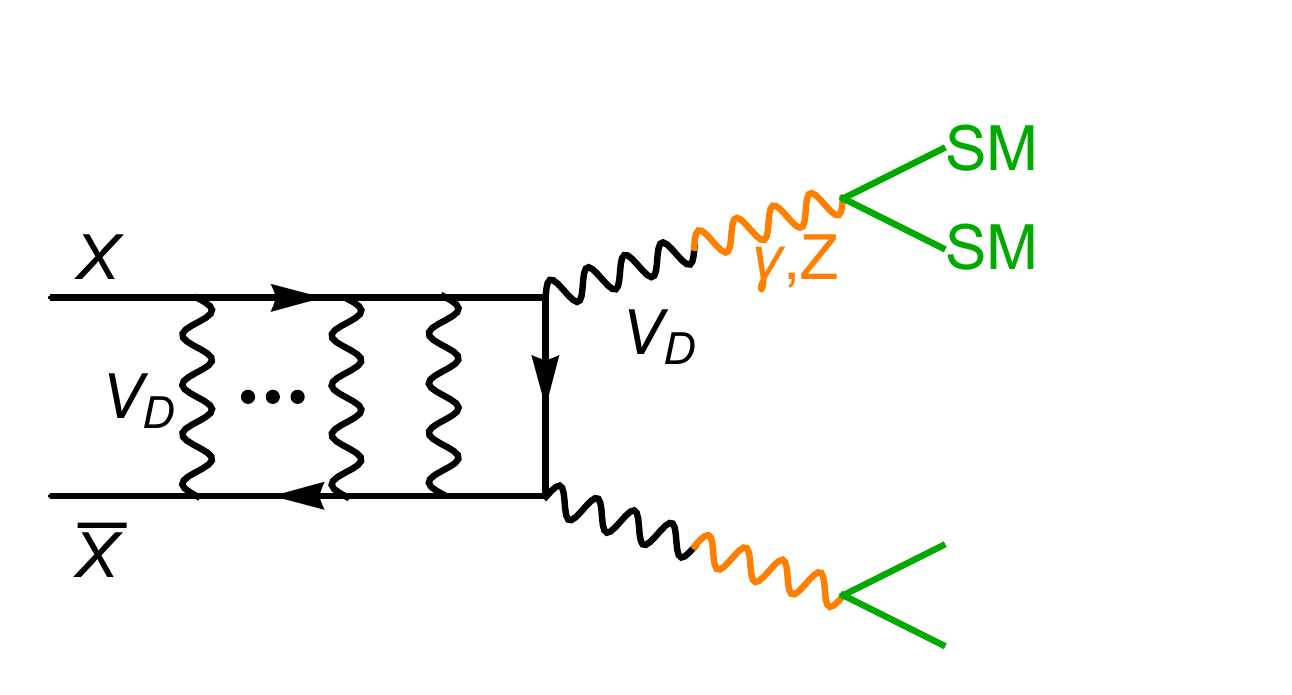}
\end{minipage}
\hspace{1cm}
\begin{minipage}{0.31 \textwidth}
\vspace{-1cm}
\includegraphics[width= 1.25 \textwidth]{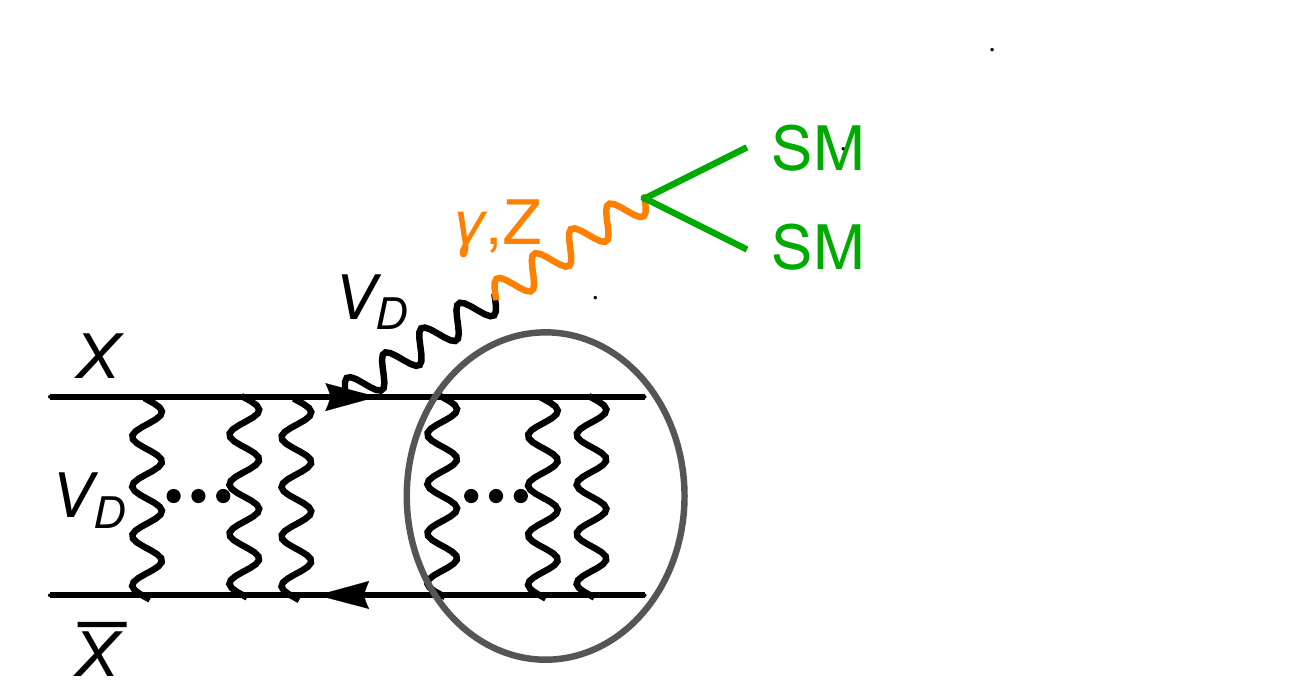}
\end{minipage}
\hspace{-0.75cm}
\begin{minipage}{0.31 \textwidth}
\includegraphics[width= 1.25 \textwidth]{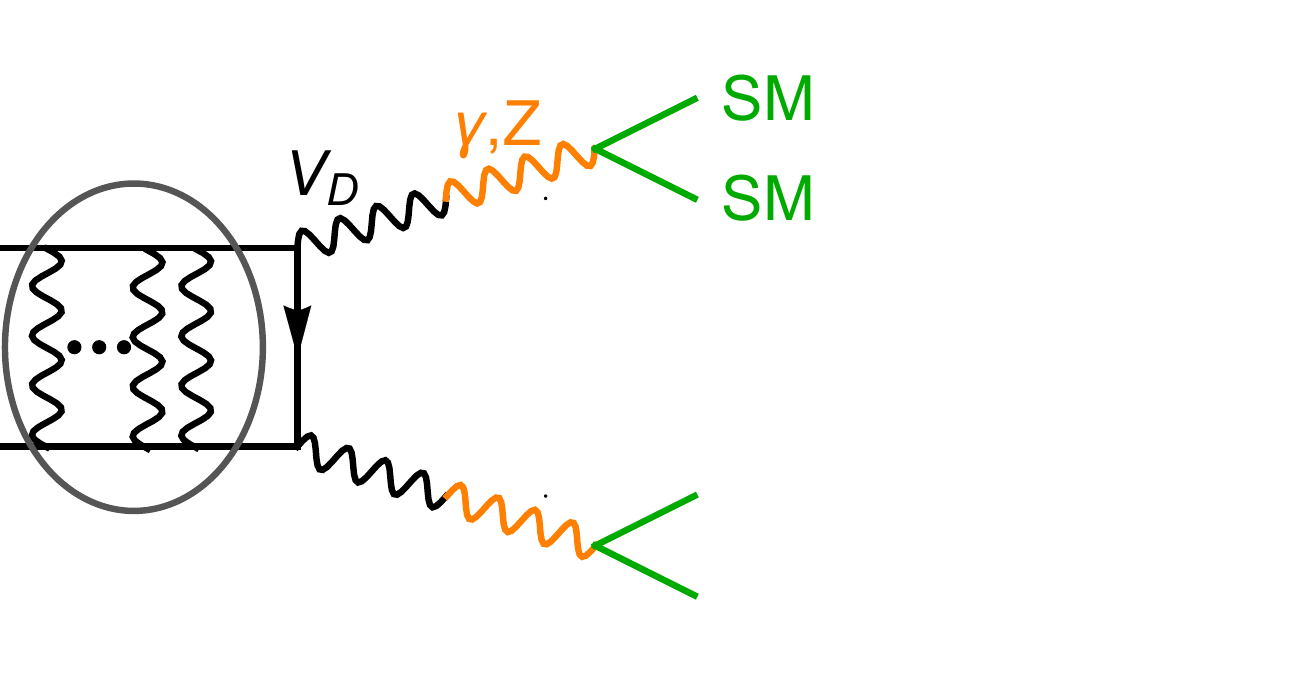}\\[-0.1cm]
\includegraphics[width= 1.25 \textwidth]{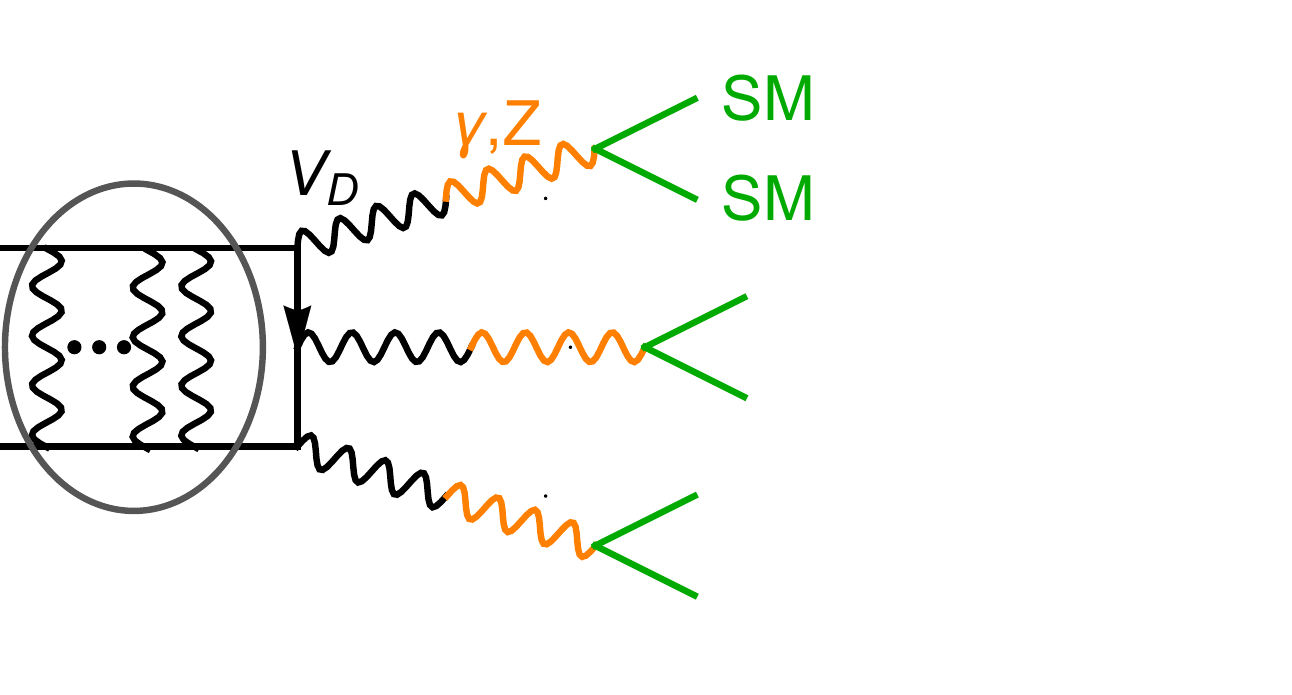}
\end{minipage}

\caption{\em \small \label{fig:BSF} {\bfseries DM annihilation} directly into dark photons (left), and via the 2-step process of {\bfseries formation and subsequent decay of bound states} (right), for the case of para- and ortho- configurations (top and bottom diagrams, respectively). The dark photons $V_D$ are produced on-shell and decay into SM particles via their kinetic mixing to the hypercharge.}
\end{figure}

Dark matter annihilation proceeds either directly into a pair of dark photons (\cref{fig:BSF} left)
\beq
X + \bar{X} \ \to \ 2 \, \VD \, ,
\label{eq:ann}
\eeq
or, in a two-step process, through the radiative formation of particle-antiparticle bound states,
\begin{equation}
X + \bar{X} \ \to \ {\cal B}_s(X\bar{X})+ \VD \, ,
\label{eq:BSF}
\eeq
and the subsequent decay of the latter (\cref{fig:BSF} middle and right)
\begin{subequations}
\label{eqs:BS decay}
\begin{align} 
{\cal B}_{_{\uparrow\downarrow}} (X\bar{X}) \ &\to \ 2 \, \VD \, ,
\label{eq:para decay}
\\
{\cal B}_{_{\uparrow\uparrow}} (X\bar{X}) \ &\to \ 3 \, \VD \, .
\label{eq:ortho decay}
\end{align}
\end{subequations}
In the interaction~\eqref{eq:BSF}, the subscript $s$ denotes the spin of the bound state that forms.
Spin-singlet (para-) states, denoted by $\uparrow\downarrow$, form 25\% of the time; spin-triplet (ortho-) states, denoted by $\uparrow\uparrow$, form 75\% of the time.\footnote{
In the non-relativistic regime, and to lowest order in the coupling $\alphaD$, the spin of each of the interacting particles remains unchanged by the capture process into a bound state. The BSF cross-section is the same for any initial spin configuration, with the final spin configuration mandated to be the identical to the initial one. For an unpolarised ensemble of unbound particles, the probability of capture into the singlet or triplet spin states thus depends solely on their multiplicities.} 
The spin of the bound state determines its dominant decay mode, namely into 2 $V_D$ or 3 $V_D$, as seen in~\eqref{eqs:BS decay}. The decay rates of the bound states read $\Gamma_{\uparrow \downarrow} = \alphaD^5 \MDM/2$ and $\Gamma_{\uparrow \uparrow} = [4(\pi^2-9)/(9\pi)] \, \alphaD^6 \MDM/2$, so that within the parameter space of interest the decay lengths range from $\sim 10^{-4}$~pc to many orders of magnitude below. Since this is rather prompt in astrophysical scales, the rate of radiative signals produced depends solely on the annihilation and BSF cross-sections. We discuss these cross-sections in \cref{sec:cross-sections}.

The dark photons $\VD$ produced in the processes \eqref{eq:ann} -- \eqref{eqs:BS decay} decay into SM particles via the kinetic mixing $\epsilon$, with the decay rate and branching ratios that we discuss in detail in \cref{sec:BRs}. While the dark photons produced in \eqref{eq:ann} and \eqref{eqs:BS decay} carry energy of the order of the DM mass, the dark photon produced in the BSF process~\eqref{eq:BSF} carries away only the binding energy and the kinetic energy of the $X-\bar{X}$ relative motion, $\omega \simeq (\MDM/4) (\alphaD^2 + v_{\rm rel}^2) \ll \MDM$, where $v_{\rm rel}$ is the relative velocity of the interacting pair in units of the speed of light $c$. Because of the larger astrophysical backgrounds at low energies, the constraints we derive in \cref{sec:DMID} emanate solely from the decay products of the high-energy dark photons emitted in \eqref{eq:ann} and \eqref{eqs:BS decay}.

\subsection{Annihilation and bound-state formation cross-sections \label{sec:cross-sections}}

\begin{figure}[t]
\begin{center}
\includegraphics[width=0.6 \textwidth]{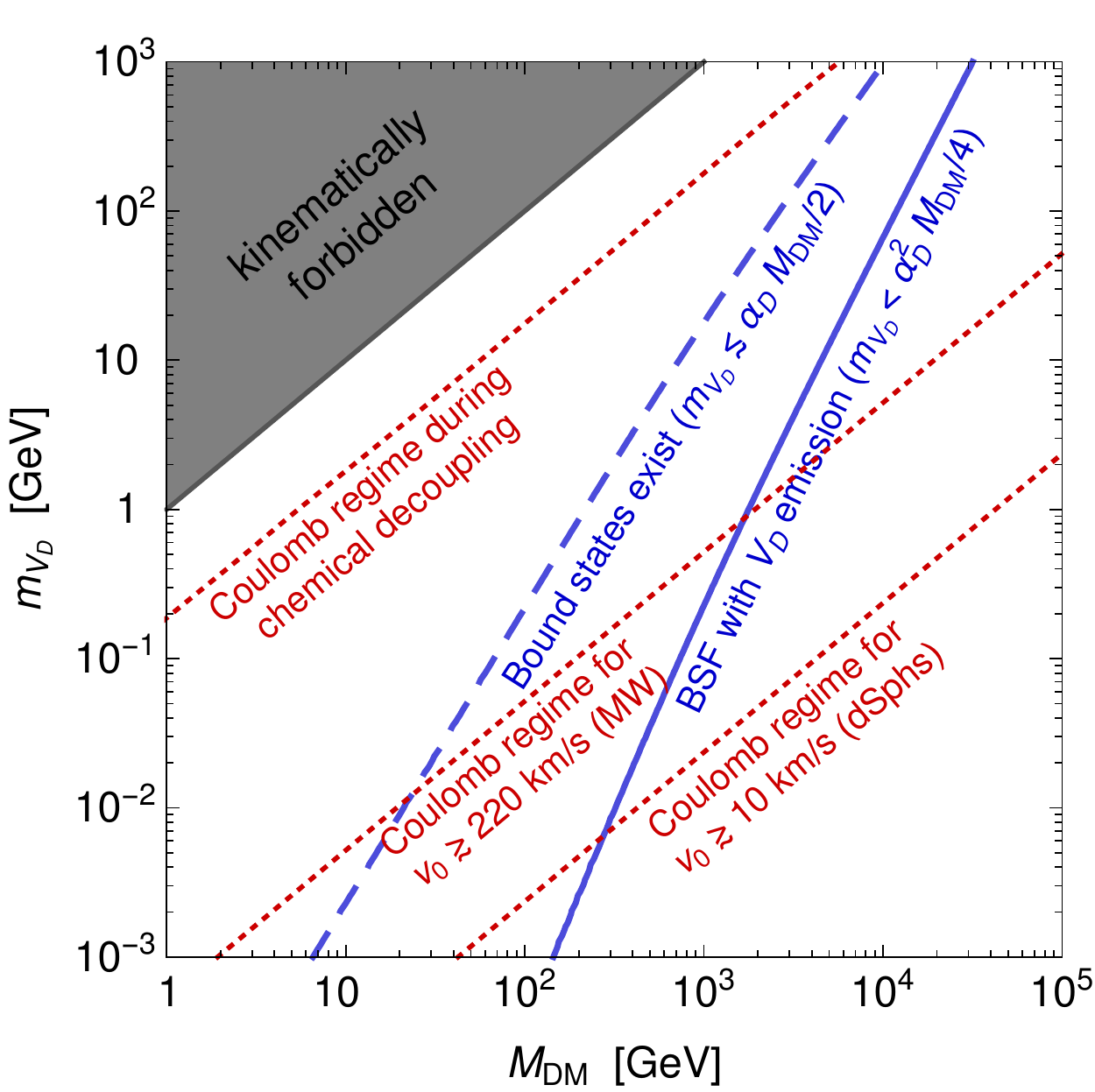}
\end{center}
\caption[]{\em \small \label{fig:PhaseSpace} The {\bfseries parameter space} considered in this work. The dark fine structure constant $\alphaD$ is determined on this plane, by requiring that the observed DM abundance arises from thermal freeze-out in the dark sector. 
Below the dashed blue line, the Bohr momentum is larger than the mediator mass, and the interaction manifests as long-range: bound states exist and the Sommerfeld effect is significant. However, bound states can form radiatively only below the solid blue line, where the binding energy is sufficient for a massive dark photon to be emitted. 
Below the red dotted lines, the average momentum transfer is larger than the mediator mass, $(\MDM/2) v_{\rm rel} \gtrsim \mVD$, for typical velocities in Dwarf galaxies, the Milky Way and during chemical decoupling of DM from the dark photon bath in the early universe. (We use  $v_{\rm rel} = \sqrt{2}v_0$.) In this regime, the Coulomb approximation for the annihilation and BSF cross-sections is satisfactory.}
\end{figure}

Bound states may form if the following conditions are met.
\begin{enumerate}[(i)]
\item 
The interaction carried by $\VD$ is effectively long-range, i.e. the inverse mass of the carrier is larger than the Bohr radius of the bound state
\begin{subequations}
\begin{equation}
\frac{\alphaD \, \MDM}{2\mVD} \gtrsim 0.84 \,.
\label{eq:longrange}
\end{equation}
This condition ensures the existence of at least one bound state, but also denotes the regime where the Sommerfeld effect is important.

\item 
The binding energy of the newly formed state is large enough to provide the mass of the emitted $\VD$, 
\begin{equation}
\frac{\alphaD^2 \, \MDM}{4 \, \mVD} \gtrsim 1 \,.
\label{eq:BSFcondition}
\end{equation}
\end{subequations}
Here we neglect the kinetic energy of the interacting particles, since $v_{\rm rel} \lesssim \alphaD$ wherever BSF is significant. The condition~\eqref{eq:BSFcondition} is always more stringent than \eqref{eq:longrange}. 
\end{enumerate}
In \cref{fig:PhaseSpace}, we present the parameter space we focus on, and depict the conditions \eqref{eq:longrange} and \eqref{eq:BSFcondition}.

For our purposes, we shall consider only capture into the ground state, with principal and angular quantum numbers $\{n\ell m\}=\{100\}$. Contributions from capture into higher-energy levels may give mild enhancements to the DM radiative signals, but are mostly subdominant either with respect to capture to the ground state, or with respect to annihilation, as explained in the discussion that follows. Moreover, the radiative capture to $n>1$ bound states is kinematically possible for a more limited range of $\mVD$ than that defined by the condition~\eqref{eq:BSFcondition}, by roughly a factor of $n^2$. In the following, any BSF cross-section presented in equations or graphs corresponds to capture to the ground state, unless otherwise stated.

\bigskip

The (spin-averaged) annihilation and BSF cross-sections can be expressed as follows~\cite{Petraki:2016cnz}
\begin{subequations}
\label{eqs:sigmas}
\begin{align}
\sigma_{\rm ann} v_{\rm rel} &= \sigma_0 \, S_{\rm ann} \,,
\label{eq:sigma_ann}
\\
\sigma_{\rm BSF} v_{\rm rel} &= \sigma_0 \, S_{\rm BSF} 
\times \pss^{1/2} (3-\pss)/2
\,,
\label{eq:sigma_BSF}
\end{align}
\end{subequations}
where $\sigma_0 \equiv \pi \alphaD^2/\MDM^2$ is the perturbative value of the annihilation cross-section times relative velocity, and 
\beq 
\pss \equiv 1- 16 \mVD^2/ [\MDM(\alphaD^2 + v_{\rm rel}^2)]^2 
\label{eq:pss}
\eeq 
is the phase-space suppression due to the emission of a massive dark photon in the capture process. 
Note that the factor $3-\pss$ accounts for the contribution from the longitudinal dark photon polarisation. For a massless dark photon, $\pss^{1/2} (3-\pss)/2 =1$.

In the Coulomb limit of a massless dark photon, $S_{\rm ann}$ and $S_{\rm BSF}$ depend only on the ratio $\alphaD/v_{\rm rel}$, and can be computed analytically (see e.g.~\cite{Petraki:2015hla}), 
\begin{subequations}
\label{eqs:Sfactors_Coulomb}
\begin{align}
S_{\rm ann}^C &= \frac{2\pi \alphaD/v_{\rm rel}}{1-e^{-2\pi \alphaD/v_{\rm rel}}} \,,
\label{eq:S_ann_C}
\\
S_{\rm BSF}^C &= \frac{2\pi \alphaD/v_{\rm rel}}{1-e^{-2\pi \alphaD/v_{\rm rel}}} 
\ \frac{(\alphaD/v_{\rm rel})^4}{[1+(\alphaD/v_{\rm rel})^2]^2} 
\ \frac{2^9}{3} \ e^{-4(\alphaD/v_{\rm rel}) \: {\rm arccot} (\alphaD/v_{\rm rel})} \,.
\label{eq:S_BSF_C}
\end{align}
\end{subequations}
For $v_{\rm rel} > \alphaD$, BSF is very suppressed. However, in the regime where the Sommerfeld effect is important, $v_{\rm rel} \lesssim \alphaD$, both the annihilation and BSF cross-sections exhibit the same velocity dependence, $\sigma v_{\rm rel} \propto 1/v_{\rm rel}$, with BSF being the dominant inelastic process, $\sigma_{\rm BSF} / \sigma_{\rm ann} \simeq 3.13$.\footnote{
\label{foot:BSF_n=2,l=1_Coulomb}
In this regime, the capture into $n=2, \ell=1$ bound states, if kinematically allowed, is also somewhat faster than annihilation~\cite{Petraki:2016cnz}. However, it is subdominant with respect to the capture to the ground state, and we ignore it in our analysis.} 
The Coulomb limit of eqs.~\eqref{eqs:Sfactors_Coulomb} is a satisfactory approximation provided that the average momentum transfer exceeds the mediator mass~\cite{Petraki:2016cnz}, 
\beq
(\MDM/2) v_{\rm rel} \gtrsim \mVD \,. \label{eq:CoulombCondition}
\eeq 
Note that this condition may hold, even if the phase-space suppression $\pss$ in the case of BSF implies that the latter is suppressed or entirely disallowed.

For $\mVD>0$, $S_{\rm ann}$ and $S_{\rm BSF}$ cannot be computed analytically. A detailed computation of the radiative BSF cross-sections for DM interacting via a Yukawa potential, and comparison with annihilation has been recently performed in  ref.~\cite{Petraki:2016cnz} (see also \cite{An:2016gad}). Here, we summarise some features that are important for the DM phenomenology.
\paragraph{Resonances.} Both the annihilation and the BSF cross-sections exhibit resonances,
which emanate from the scattering (initial) state wavefunction, and occur at discrete values of the ratio of the Bohr momentum to the mediator mass, $\alphaD \MDM/ (2\mVD)$. These values correspond to the thresholds (maximum $\mVD$ or minimum $\alphaD \MDM$) for the existence of bound-state levels. Since for a Yukawa potential, the bound-state energy levels depend both on the principal and the angular quantum numbers, $n$ and $\ell$, so do the thresholds for their existence. Thus, each $\ell$-mode of the scattering state wavefunction exhibits a different spectrum of resonances that corresponds to the thresholds for the existence of bound-state levels of the same $\ell$. 

\begin{figure}[!t]
\begin{center}
~\includegraphics[width= 0.48 \textwidth]{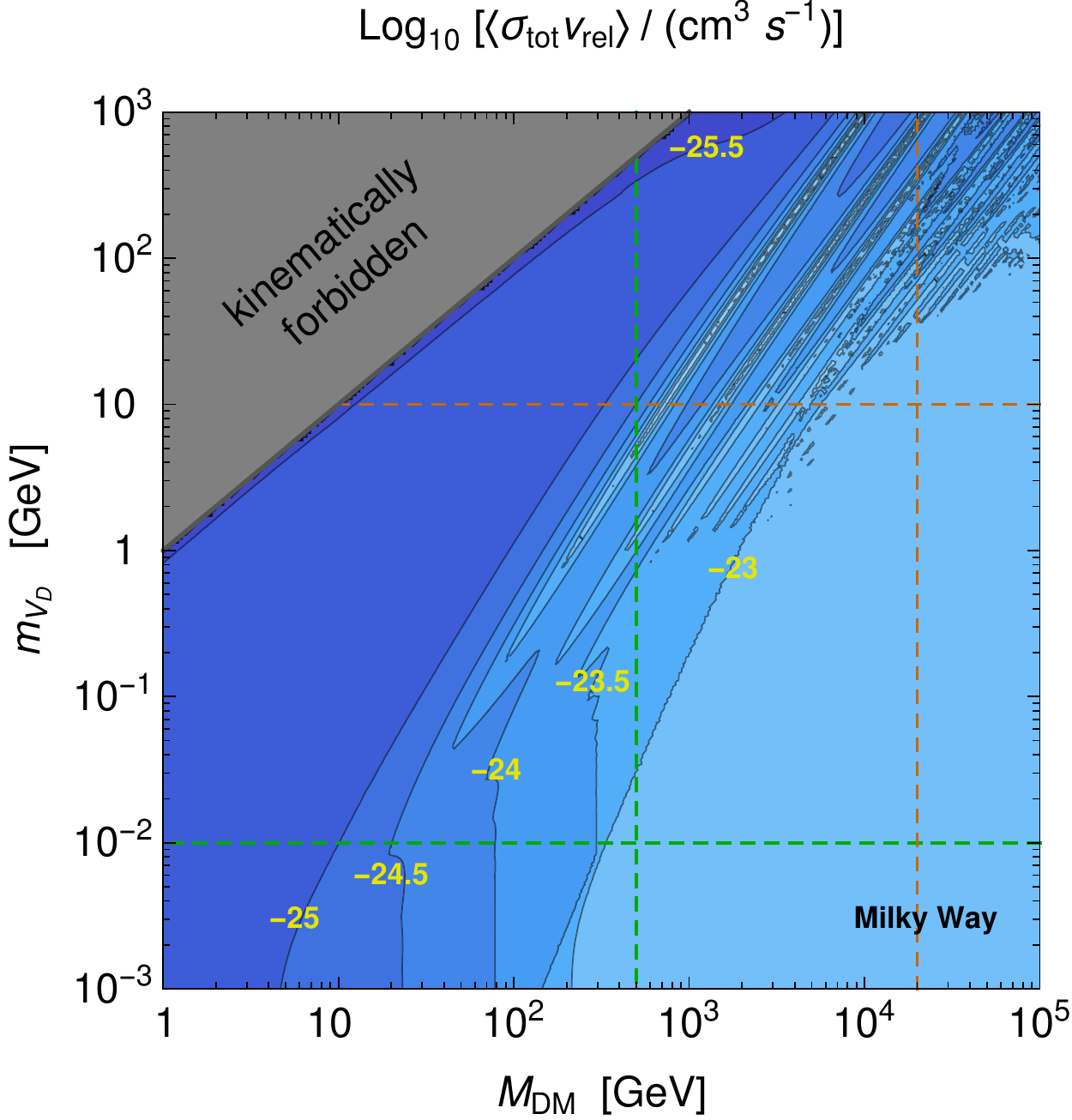}~
\includegraphics[width= 0.485 \textwidth]{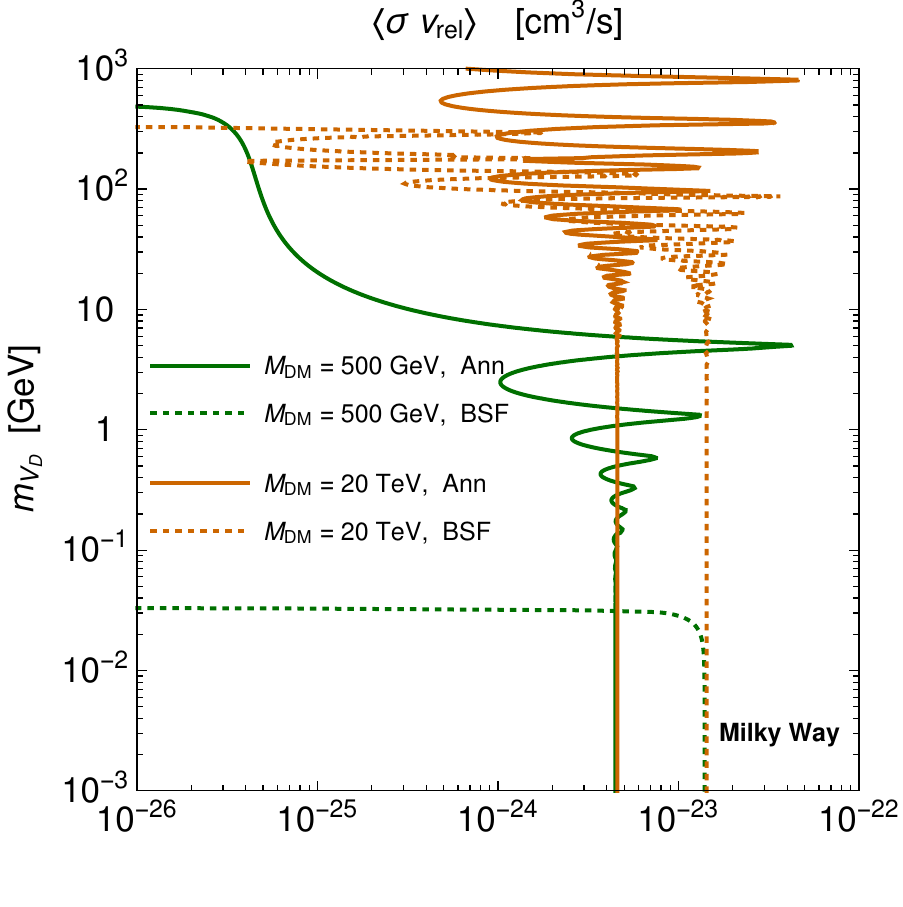}
\\
\includegraphics[width= 0.485 \textwidth]{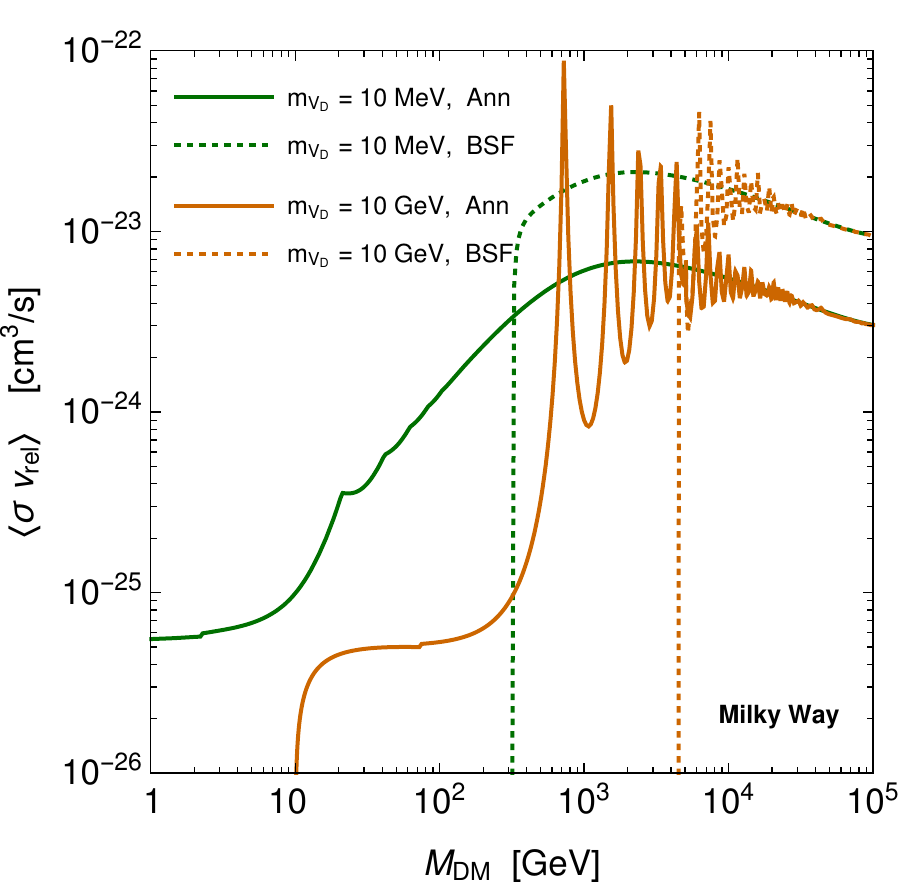}~
\includegraphics[width= 0.48 \textwidth]{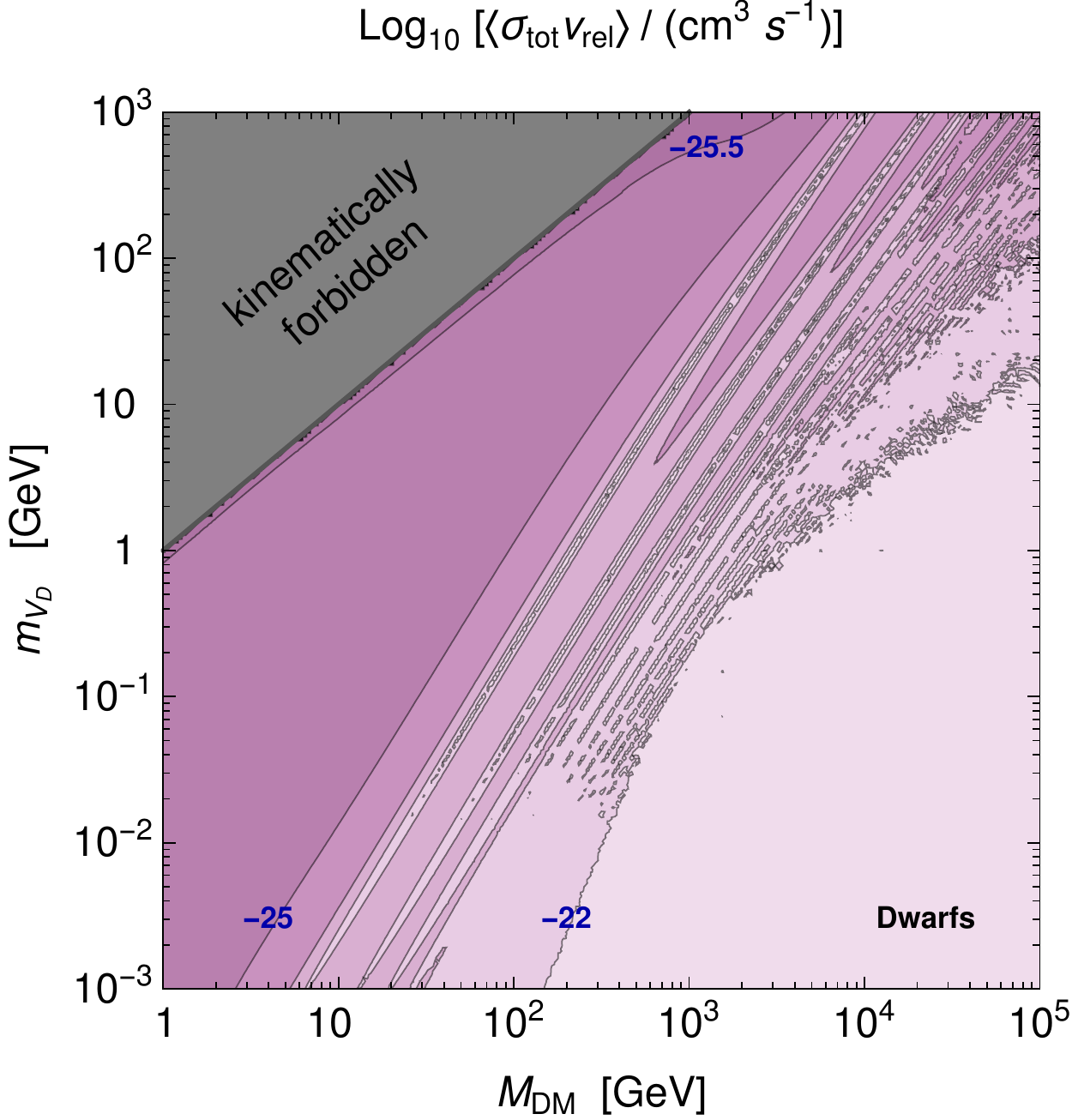}~
\end{center}
\caption{\em \small \label{fig:CrossSections_MWandDSphs}
{\bfseries Velocity averaged cross-sections}. The top-left panel shows the value of the total cross-section in the Milky Way (MW), in the entire parameter space that we consider. The bottom-right panel shows the same quantity in Dwarf spheroidal galaxies. The bottom-left and top-right panels show slices of the MW cross-section, as a function of $\MDM$ for specific values of $\mVD$ (bottom-left) and vice-versa (top-right). These panels show explicitly the separate contributions to the cross-section of annihilations (solid lines) and BSF (dashed lines).}
\end{figure}

Angular momentum selection rules imply that (to leading order in $\alphaD$ and $v_{\rm rel}$) the $\ell$-modes of the scattering state wavefunction that participate in the two processes of interest, are different. 
The DM annihilation into two vector bosons \eqref{eq:ann} receives its dominant contribution from the $\ell=0$ mode. On the other hand, the radiative formation of the ground state \eqref{eq:BSF}, receives its dominant contribution from the $\ell=1$ mode; this orbital angular momentum of the initial state provides for the spin of the vector boson emitted in the capture process into the ground state (which has zero orbital angular momentum). Thus, the DM radiative signals include both $\ell=0$ and $\ell=1$ resonances. The resonance structure of annihilation and BSF can be seen in \cref{fig:CrossSections_MWandDSphs}.

\paragraph{Velocity dependence.} Besides the resonances, the $\ell$ modes of the scattering state wavefunction that participate in a process determine the velocity dependence of the corresponding cross-section away from the Coulomb regime. At sufficiently low velocities, $\sigma v_{\rm rel} \propto v_{\rm rel}^{2\ell}$; therefore, the annihilation cross-section saturates at a velocity-independent value, while the cross-section for capture into the ground state becomes $v_{\rm rel}^2$ suppressed. 
This behaviour holds both on- and off-resonance. The departure from the Coulombic behaviour occurs at $v_{\rm rel} \lesssim 2\mVD/\MDM$. There is an important difference though, between resonant and non-resonant points. 
For non-resonant $\alphaD \MDM/ (2\mVD)$ values, the $\sigma v_{\rm rel} \propto v_{\rm rel}^{2\ell}$ behaviour is established at $v_{\rm rel} \lesssim 2m_{\VD}/\MDM$, and the annihilation and BSF cross-sections remain always below their Coulomb values at the same speed. 
In contrast, for $\alphaD \MDM/ (2\mVD)$ values near or on-resonance, the cross-sections grow above their Coulomb value at $v_{\rm rel} \lesssim 2\mVD/\MDM$, and acquire the $\sigma v_{\rm rel} \propto v_{\rm rel}^{2\ell}$ behaviour at much lower velocities.
The above imply that BSF is comparable to, or stronger than annihilation within a range of velocities that depends on  $\alphaD \MDM/ (2\mVD)$. Away from resonances, this speed range is essentially limited within the Coulomb regime and by the onset of the Sommerfeld effect ($2\mVD/\MDM \lesssim v_{\rm rel} \lesssim \alphaD$); however, it is extended to significantly lower velocities, on or near resonances~\cite{Petraki:2016cnz}.\footnote{
\label{foot:BSF_n=2,l=1_nonCoulomb}
Our discussion suggests that away from the Coulomb regime, the capture into the $n=2,\ell=1$ bound states may become the dominant inelastic process. This is because the angular momentum selection rules imply that the $\ell=0$ and $\ell=2$ modes of the scattering state wavefunction participate in this process. At low velocities, due to the $v_{\rm rel}^{2\ell}$ scaling described above, the contribution from the $\ell=0$ mode renders the capture into $n=2, \ell=1$ states more significant than the capture into the ground state. However, the contribution from this mode alone is subdominant to the annihilation. We conclude that the formation of $n=2, \ell=1$ states can be neglected, either with respect to the capture to the ground state (Coulomb limit, cf.~footnote~\ref{foot:BSF_n=2,l=1_Coulomb}), or with respect to annihilation (except perhaps very close to $\ell=2$ resonances, which however cover very limited parameter space).}

For our purposes, BSF is significant in Dwarf galaxies, in the Milky Way and during the chemical decoupling of DM in the early universe, within large portions of the parameter space where the radiative capture is kinematically possible. 
In \cref{fig:PhaseSpace}, we circumscribe the parameter space where the Coulomb regime is attained, for the corresponding typical DM speeds (see below); in these regimes, the analytical formulae~\eqref{eqs:Sfactors_Coulomb} may be used (although the phase-space factor appearing in \cref{eq:sigma_BSF} should still be properly taken into account). The transition from the Coulomb regime to the resonant regime, with increasing $\mVD$ or decreasing $\MDM$, can be seen in \cref{fig:CrossSections_MWandDSphs}.
For the range of dark photon masses considered, BSF is unimportant during CMB, when the DM average speed is extremely low (cf.~\cref{sec:CMB}). 

\medskip

Before moving on, let us specify here for reference the speed distributions we adopt. For the Milky Way and for dwarf galaxies, we consider a Maxwellian distribution with a cutoff at $v = v_{\rm esc}$,
\beq 
f(\mathbf{v}) = N(v_0,v_{\rm esc}) \ \Theta(v_{\rm esc}-v) \ e^{-v^2/v_0^2} \,.
\nonumber \eeq
The normalization factor $N(v_0,v_{\rm esc})$ is chosen  such that $\int d^3v \, f(\mathbf{v})=1.$
For the Milky Way, we take $v_0=220$ km/s and $v_{\rm esc}=533$ km/s, which are appropriate values for the DM phase-space distribution in our galaxy (see e.g.~\cite{Piffl:2013mla}).
For dwarfs, we take $v_0=10$ km/s and $v_{\rm esc}=15$ km/s, which are values inferred from the typical velocities of stellar tracers in the dwarfs we will consider later (see e.g.~\cite{McConnachie:2012vd}), increased by a factor $\sim 1.6$ as prescribed by the empirical relations in~\cite{Burkert:2015vla}.
For the CMB epoch, see the dedicated discussion in \cref{sec:CMB}.
We note that, in the limit of $v_{\rm esc} \to \infty$, the distribution of the two-particle relative velocity is also a Maxwellian with $v_{0,\rm rel} = \sqrt{2} v_0$. For a finite $v_{\rm esc}$, the $v_{\rm rel}$ distribution is more complex. While we take this into account in our computations, the aforementioned relation between the average values of $v$ and $v_{\rm rel}$ is useful for estimations.

\subsection{Dark photon decay}
\label{sec:BRs}

The kinetic mixing with hypercharge of \cref{eq:L} implies that the dark photon mixes with both the SM photon and $Z$. This induces the following couplings between $\VD$ and the SM fermions~$f$
\beq
\mathcal{L} \supset g_{f} \VD^\mu (\bar{f} \gamma_\mu f),\qquad g_f 
= \epsilon\,e\left(Q_f\frac{1}{1-\delta^2} 
+ \frac{Y_f}{c_w^2} \frac{\delta^2}{\delta^2-1}\right) + O(\epsilon^2)\,,
\label{eq:couplings}
\eeq
where $\delta = \mVD/m_Z$, $Q_{f=e_L,e_R,u_L,\dots} = -1,-1,2/3,\,\dots$ and $Y_{f=\ell_L,\ell_R,u_L,\dots} = -1/2,-1,1/6,\,\dots$., so that the non-vectorial structure of the $g_f$ coupling is left implicit.  \Cref{eq:couplings} makes it manifest that the $\VD$ couplings to SM particles are $\epsilon$-suppressed, and that they are proportional to their electric charge for $\mVD \ll m_Z$, and to their hypercharge for $\mVD \gg m_Z$ (\textit{i.e.} when $U(1)_Y$ is unbroken). 
Of course, the $\epsilon$ expansion of \cref{eq:couplings} is not valid for $\delta \sim 1$. 
In that limit, the $\VD$ couplings depend on the $f$ quantum numbers as the $Z$ boson ones. In our study, we use the full tree-level expressions for $g_f$, that are valid also in the $m_Z \sim m_\VD$ limit. The interested reader may find them in~\cite{Babu:1997st,Curtin:2014cca}.

We compute the widths of $\VD$ at tree level\footnote{Loop corrections do not affect substantially this result, as can be seen comparing our values with the ones given in ref.~\cite{Curtin:2014cca}, where such corrections have been taken into account.} as
\beq
\Gamma(\VD \to f\bar{f}) = \frac{N_f}{24 \pi}  \mVD \sqrt{1-\frac{4 m_f^2}{\mVD^2}} \,
\left[ g_{f_L}^2 + g_{f_R}^2 - \frac{m_f^2}{\mVD^2} (g_{f_L}^2 + g_{f_R}^2 - 6 g_{f_L} g_{f_R} )\right],
\eeq
where $N_f = 1$ for $f = \ell, \nu$ and $N_f = 3$ otherwise.
We do not report the widths into $WW$ and $Zh$, that even for $\mVD > 2 m_W, m_h+m_Z$ contribute less than $\sim 10\%$ to the total $\VD$ width.
In the interval $\Lambda_{\rm QCD} < \mVD < 5$~GeV we cannot treat the light quarks as free particles, because QCD is strongly coupled and decays should be described in terms of QCD resonances. For example, in {\sc Pythia}, which is the tool we use to produce all our spectra (cf.~\cref{sec:DMID}), the decay $\tau^- \to \nu_\tau \pi^-$ is treated by an effective $\tau \nu_\tau \pi$ vertex (see \textit{e.g.}~\cite{Ilten:2012zb}). 
We choose 5 GeV as the upper extreme of the interval dominated by hadronic decays, because {\sc Pythia} has not been optimised for energies below that value, and we set the lower extreme to the reference value of 350~MeV.
As customary (see \textit{e.g.}~\cite{Curtin:2014cca,Buschmann:2015awa} for recent works), we determine the hadronic decay width from measurements of $e^+ e^- \to$ hadrons. This process is dominated by $\gamma^*$ exchange, and therefore provides a good handle to describe the hadronic decays of our spin-1 $\VD$ initial state. The related $\VD$ decay width then reads
\beq
\Gamma(\VD \to {\rm hadrons}) = R(s = \mVD^2)\,\Gamma(\VD \to \mu^+\mu^-),
\eeq
where we extract $R(s) = \sigma(e^+ e^- \to {\rm hadrons})/\sigma(e^+ e^- \to \mu^+ \mu^-)$ from~\cite{Curtin:2014cca}, with $\sqrt{s}$ center-of-mass energy of the $e^+e^-$ collision.

To our knowledge, no tool has been developed that would allow to obtain, from the hadrons produced for $\Lambda_{\rm QCD} < \mVD < 5$~GeV, the energy spectra of stable SM particles that we need in deriving our bounds (namely $\gamma$-rays, $e^\pm$, $\bar p$).
However, the description can be simplified by observing that $\pi^+\pi^-$ pairs largely dominate the final states from the decays of the resonances produced in $\VD \to$~hadrons (see \textit{e.g.}~\cite{Olive:2016xmw,Buschmann:2015awa,Whalley:2003qr}). Since BR$(\pi^\pm \to \mu^\pm \nu_\mu) > 99.9\%$, the final states from those hadrons are then dominated by $\mu^+\mu^-$ and $\nu_\mu \bar{\nu}_\mu$ pairs, up to a very small fractions of other SM particles, like photons from $\pi_0$ decays.
Therefore we employ the following approximation: we assume that half of the hadronic width consists in the $\mu^+\mu^-$ final state, and the other half in the $\nu_\mu \bar{\nu}_\mu$ one.
The resulting energy distribution of final muons is then less broad, and peaked at somehow larger energies, than the one that would arise from following properly all the steps of the various hadronic cascades (see \textit{e.g.} ref~\cite{Elor:2015bho}, which discusses the analogous case of multistep cascades of a hidden sector).
We believe our approximation to be nonetheless sufficient for our purposes. On one hand, the interval of $\mVD$ in which these subtleties are relevant is only a portion of the much larger range which we consider. On the other hand,  even within this interval, 
i) the CMB constraints are not affected, because they are not sensitive to the number of steps of a cascade~\cite{Elor:2015bho}; 
ii) our constraints may be taken as possibly slightly aggressive for $\MDM \lesssim 100$~GeV, since enhancing the lower energy $\gamma$'s from $\mu$ showers, at the expense of higher energy $\gamma$'s,
would typically bring the resulting spectrum below the Fermi sensitivity; 
iii) the treatment chosen affects mildly a region that is anyway excluded by CMB.

\medskip

To summarise, we compute the $\VD$ branching ratios to SM particles at tree level outside the `hadronic decays' interval $350~{\rm MeV} < \mVD < 5$ GeV, where instead we extract the light hadrons width from measurements of $e^+e^- \to $ hadrons at colliders. We then assume this width to consist in half $\mu^+\mu^-$ and half $\nu_\mu \bar{\nu}_\mu$ pairs. The resulting BRs, that we use in the rest of this paper, are displayed in \cref{fig:DPBR}.

\begin{figure}[t]
\begin{center}
\includegraphics[width= 0.48 \textwidth]{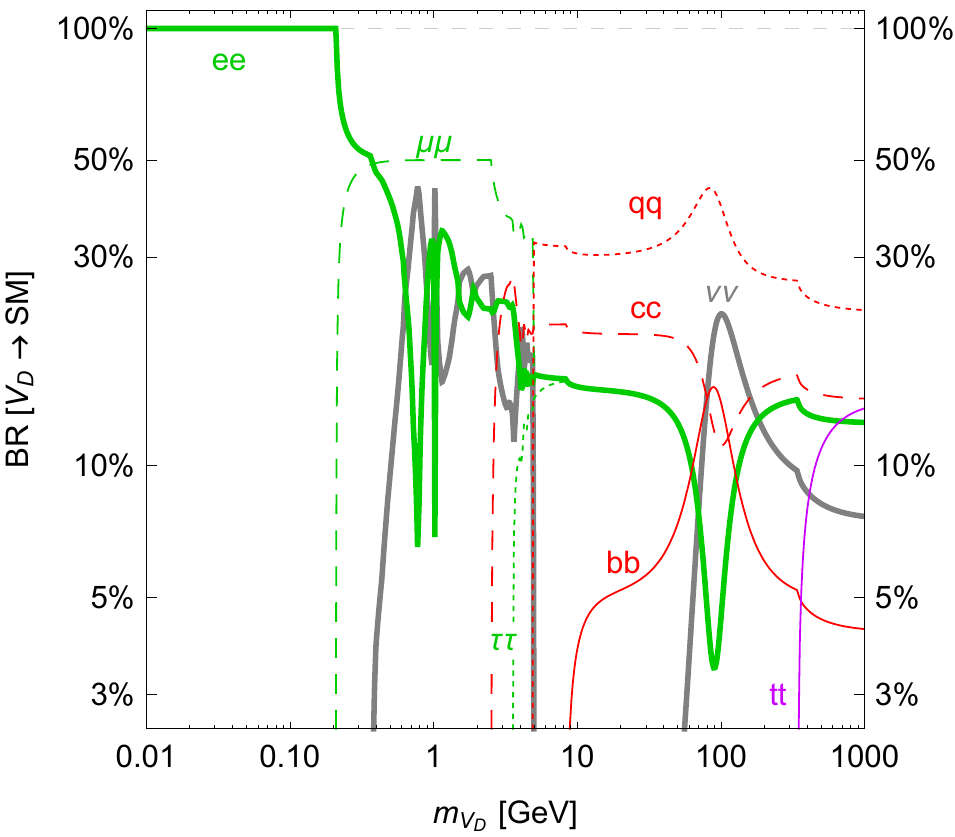}
\caption{\em \small \label{fig:DPBR} \label{fig:DDbounds} 
{\bfseries Branching ratios} of the dark photon into SM fermions. }
\end{center}
\end{figure}

\subsection{Constraints on the kinetic mixing \label{sec:epsilon_constraints}}

\subsubsection{Supernovae and beam dump experiments}
\label{sec:SNbeamdump}

We summarize here the constraints on the kinetic mixing $\epsilon$ that are independent of the existence of any dark sector state other than $\VD$, focusing on the region $\epsilon \lesssim 10^{-3}$. We refer the reader to~\cite{Essig:2013lka,Alexander:2016aln} for recent reviews.

\paragraph{Supernovae (SN).}
Light dark photons with a mixing parameter in the approximate range $\epsilon \simeq 10^{-10}-10^{-6}$ are constrained by the neutrino spectrum observed~\cite{Hirata:1987hu,Bionta:1987qt,Alekseev:1987ej} after the explosion of supernova SN1987A, see~\cite{Dent:2012mx, Kazanas:2014mca, Rrapaj:2015wgs, Chang:2016ntp, Hardy:2016kme} for recent works.
According to the standard argument~\cite{Raffelt:1996wa}, the cooling of the SN core becomes  more efficient if enough SM photons from the explosion oscillate into $\VD$ (which requires a minimum value of $\epsilon$), 
and if enough $\VD$ escape the supernova without further interacting nor decaying (which implies a maximum value of $\epsilon$). Supernovae observations can thus exclude a finite band of $\epsilon$ values. The most updated analyses have been carried out in ref.~\cite{Chang:2016ntp,Hardy:2016kme}, where the plasma effects of finite temperature and density have been included for the first time. The results of the two analyses appear to be in reasonable agreement, and for definiteness we show the fiducial exclusion from~\cite{Chang:2016ntp} as a red shaded region in \cref{fig:eps_mVD}.
To give a sense of possible theoretical errors, we also show with a thicker red line the contours of the region that the authors of~\cite{Chang:2016ntp} name `robustly excluded', and that they find to be disallowed irrespectively of the details of their modelling.
Finally, we show in thin red the contour of the region that is excluded by the analysis of ref.~\cite{Kazanas:2014mca}, from the non-observation of a photon excess at the SMM Gamma-Ray Spectrometer~\cite{Chupp:1989kx}, that would be caused by the dark photons escaping the inner part of the SN. Since the latter bounds do not include potentially important effects~\cite{Chang:2016ntp,Hardy:2016kme} we do not shade the related area, in order to be conservative.

\paragraph{Beam dump experiments.} 
They consist of a high-intensity beam of particles sent on a fixed target, and a detector placed somewhat far from the target. They can constrain dark photons in the MeV to GeV range, for $\epsilon \sim 10^{-8}-10^{-2}$. 
When an electron beam hits the target, dark photons can be produced via bremsstrahlung, pass a shield that screens the SM background, and be detected via the leptons they decay into. The events expected from the dark photons have been computed in refs.~\cite{Bjorken:2009mm,Andreas:2012mt}, and the limits derived from the E137~\cite{Bjorken:1988as}, Orsay~\cite{Davier:1989wz}, E141~\cite{Riordan:1987aw} and E774~\cite{Bross:1989mp} data produce some of the gray shaded area in \cref{fig:eps_mVD}.
When a proton beam hits the target, dark photons can be produced both directly from the proton-target scattering, and from the decays of hadrons produced in the same scattering, as originally envisioned in ref.~\cite{Fayet:1980rr}. Detectors can then collect leptons from the decays of those dark photons that have travelled long enough. Neutrino experiments fit into this category~\cite{Batell:2009di}.
Among them, the reinterpretation of {\sc Lsnd}~\cite{Auerbach:2001wg, deNiverville:2011it,Kahn:2014sra}, {\sc Charm}~\cite{Bergsma:1985is,Gninenko:2012eq} and $\nu$-{\sc Cal I} (at the U70 accelerator)~\cite{Blumlein:1990ay,Blumlein:2011mv} data contributes to the excluded gray shaded area in \cref{fig:eps_mVD}.

\subsubsection{Direct Detection \label{sec:DD}}

Direct detection aims at revealing the tiny nuclear recoils produced by DM particles scattering off target nuclei in underground experiments.  In the non-relativistic limit, the differential cross section can be computed from the interaction Lagrangian in \cref{eq:couplings}. For $\delta \rightarrow 0$,\footnote{
This is well justified, since as one can see in the right-panel of Fig.~\ref{fig:eps_mVD}, the dark photon mass probed by direct detection experiments is much smaller than $m_Z$ for the small values of $\epsilon$ we are intererested in.} 
it reads~\cite{Fornengo:2011sz, Kaplinghat:2013yxa}
\beq\label{eq:ScatteringXS}
\frac{{\rm d}\sigma}{{\rm d}E_{\rm R}}(v,E_{\rm R}) 
= \frac{8\pi \, \alpha_{\rm {em}} \alphaD \epsilon^2 \, m_{\rm T }}{\left(2 m_{\rm T } E_{\rm R}+\mVD^2\right)^2} \frac1{v^2}Z_{\rm T}^2 F_{\rm Helm}^2(2 m_{\rm T} E_{\rm R}) \ ,
\eeq
where $m_{\rm T}$ and $Z_{\rm T} e$ are the mass and the electric charge of the target nucleus T respectively, and $\alpha_{\rm em}=e^2/4\pi$ is the fine structure constant. Here, $F_{\rm Helm}$ is the Helm form factor~\cite{Helm:1956zz,Lewin:1995rx} related to the charge density of the nucleus.

As is evident, \cref{eq:ScatteringXS} exhibits two regimes:
$i)$ for $\mVD \gg 2m_{\rm T} E_{\rm R}$, the interaction is contact-type, and for a fixed value of $\MDM$ the differential cross section scales like $\epsilon^2 /m_{V_{D}}^4$;
$ii)$ for $\mVD \ll 2m_{\rm T} E_{\rm R}$, the interaction manifests as long-range, and the cross section is independent of $\mVD$. For a given $\MDM$, the differential cross section is simply proportional to $\epsilon^2$. 
Considering typical target nuclei ($m_{\rm T}\simeq$ 100 GeV) and recoil energies ($E_{\rm R}\simeq$ 1 keV) in direct searches, the transition regime occurs for $\mVD\simeq 15$ MeV. Recent analysis of the phenomenology in direct detection of DM models with  long-range DM-nucleus interactions can be found  e.g.~in \cite{DelNobile:2015uua,DelNobile:2015bqo,Panci:2014gga,Li:2014vza}.

\begin{figure}[t]
\begin{center}
\includegraphics[height= 0.48 \textwidth]{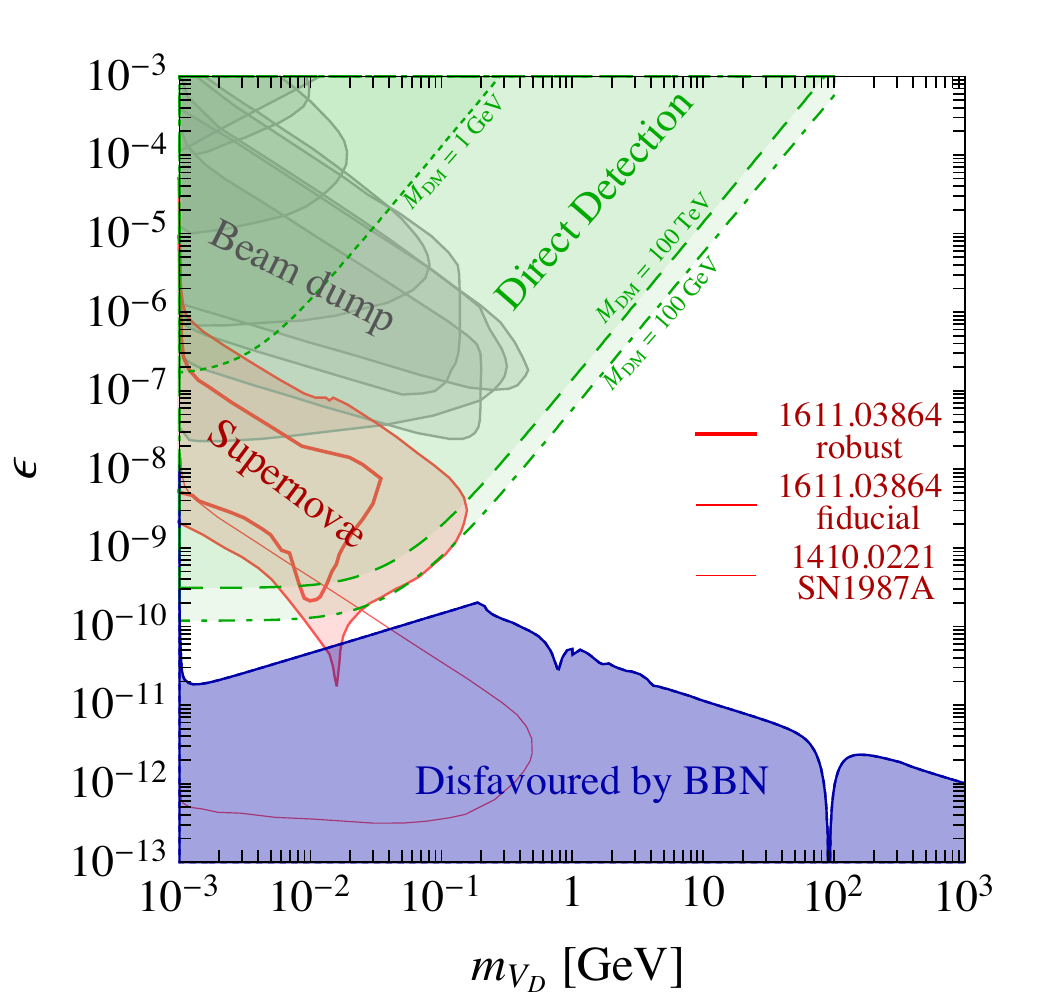}~~
\includegraphics[height= 0.48 \textwidth]{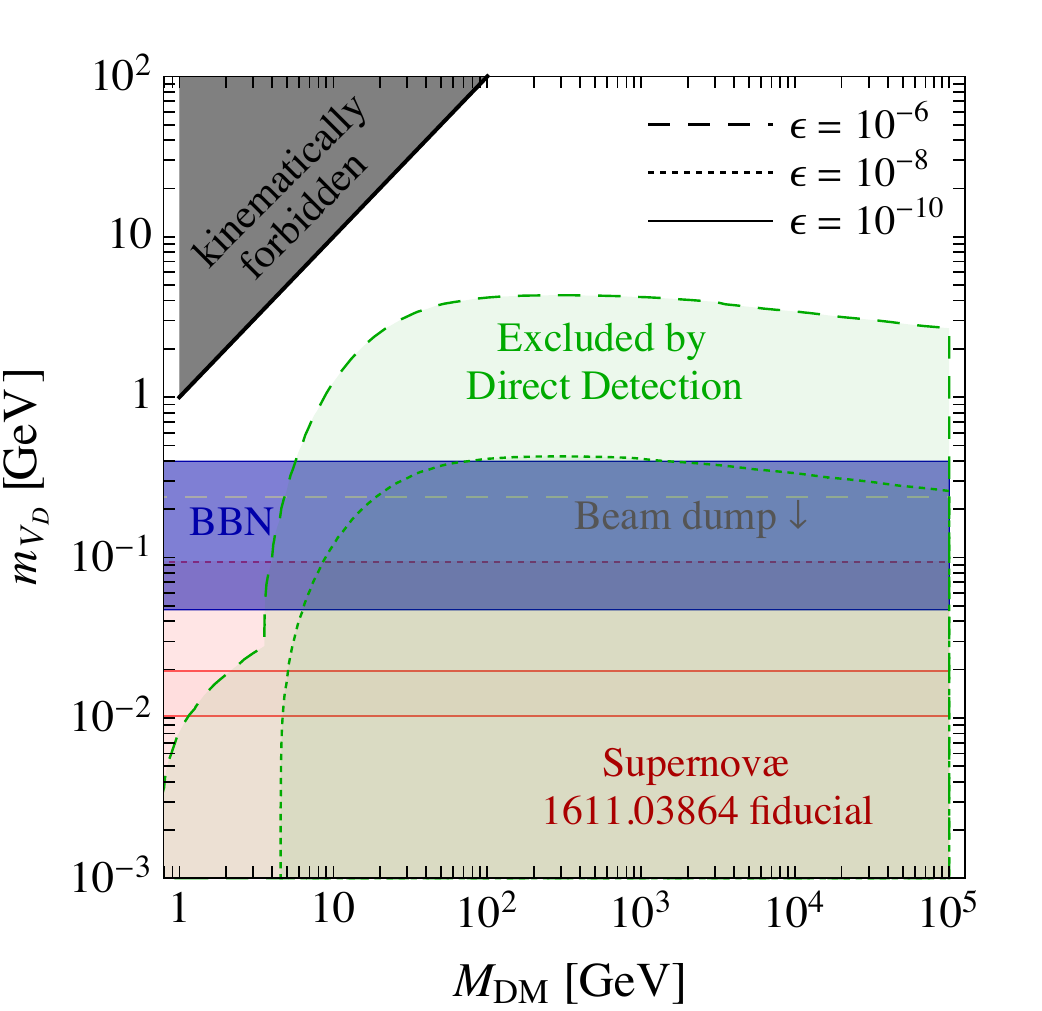}
\caption[]{\em \small \label{fig:eps_mVD} {\bfseries Constraints on the kinetic mixing}. Left: excluded regions in the plane ($\mVD,\epsilon$), taking into account several beam dump experiments (gray shaded areas), supernov\ae\ (red shaded areas), direct detection (green shaded areas, for different indicative values of the DM mass) and  BBN arguments (blue shaded area). See the text for more details. Right: the same excluded regions illustrated in the  ($\MDM,\mVD$) plane, for three selected values of the mixing parameter $\epsilon$.}
\end{center}
\end{figure}

Having at our disposal the scattering cross section, the rate of nuclear recoils, expressed in cpd (counts per day) per kilogram per keV, is then
\beq\label{eq:RateDD}
\frac{{\rm d}R_{\rm T}}{{\rm d}E_{\rm R}}= \frac{\xi_{\rm T}}{{m_{\rm T}}}\frac{\rho_\odot}{\MDM}\int_{v_{\rm min}(E_{\rm R})}^{v_{\rm esc}} \hspace{-.4cm}{\rm d}^3 v \, v f_{\rm E}(\vec v)\frac{{\rm d}\sigma}{{\rm d}E_{\rm R}}(v,E_{\rm R})  \ ,
\eeq
where $\xi_{\rm T}$ is the mass fraction of different nuclides used in direct detection, $\rho_\odot =0.3$ GeV/cm$^3$ is the local DM energy density and $f_{\rm E}(\vec v)$ is the DM speed distribution in the Earth's frame. In the velocity integral, $v_{\rm min}(E_{\rm R})$ is the minimal  speed for which DM particles can provide a given  recoil energy $E_{\rm R}$. $v_{\rm esc}$ refers of course to the Milky Way's escape speed.

\smallskip

Since the cross section in \cref{eq:ScatteringXS} is independent of the nucleus spin, we derive bounds on the relevant parameter space of the model considering the most sensitive detectors for spin-independent DM-nucleus scattering. More specifically, for $\MDM\gtrsim$ 3 GeV we use the latest \LUX\  WS2014-16 run~\cite{Akerib:2016vxi} to set limits, while for $\MDM\lesssim$ 3 GeV we derive bounds from the \CRESST\ experiment~\cite{Angloher:2015ewa}.  We follow the  procedure  of~\cite{Kavanagh:2016pyr} described in Appendix~B.2 (\LUX) and B.3 (\CRESST), where the reader can find more details.

\medskip

\Cref{fig:eps_mVD} shows, in green shaded areas, the excluded regions of the parameter space. In the left panel they are reported in the $(\mVD, \epsilon)$ plane, with the different lines (dotted, dashed-dotted, dashed) corresponding to $\MDM=1,~10^2,~10^5$~GeV respectively. In particular, for $\MDM=1$ GeV the exclusion is given by the \CRESST\ detector, while for $\MDM\gtrsim 3$ GeV is dominated by \LUX. 
For light dark photons,  the constraints become independent of $\mVD$.
As pointed out above, this is because the interaction is entering the long-range regime. For \LUX\ this transition  occurs for $\mVD \simeq 20$ MeV, while for \CRESST\ around 1 MeV, since the detector is composed by lighter target nuclei (oxygen and calcium) and employs a lower energy threshold  (around 300 eV) compared to xenon-based experiments, such as \LUX. 

In the right panel of the same figure, we project the exclusion regions in the ($\MDM,\mVD$) plane for three fixed values of $\epsilon$. 
The different lines (dashed, dotted, solid) correspond to $\epsilon = 10^{-6}, 10^{-8}, 10^{-10}$. For $\epsilon \lesssim 10^{-7}$  the bounds come solely from  \LUX ,  since the \CRESST\ detector is not able  to probe this region of the parameter space, mainly due to its lower exposure with respect to \LUX. For larger $\epsilon$, \CRESST\ becomes relevant for light DM ($\MDM \lesssim 3$~GeV).

\subsubsection{Summary \label{sec:Summary}}

We have illustrated the constraints of this section in \cref{fig:eps_mVD}, adding for completeness the region disfavoured by BBN and discussed in \cref{sec:BBN}. 
We see in the left panel that, for $m_\VD \gtrsim 1$ GeV, a wide region is still unexplored by current measurements, and an upper limit on $\epsilon$ is provided only by direct detection experiments (in the $\epsilon$ region we are considering). 
Lower dark photon masses are more constrained, and only narrow values of $\epsilon$ are allowed. 
We notice however that, for $\epsilon \sim 10^{-(10-12)}$, a more careful study of BBN constraints (and perhaps of SN bounds) would be needed to make any conclusive statement about the low $m_\VD$ region (see the discussion in~\cref{sec:BBN}).
Finally, the same constraints are shown in the right-hand plot of \cref{fig:eps_mVD} on the ($\MDM,\mVD$) plane, for three fixed values of $\epsilon.$

\section{Cosmology \label{sec:cosmo}}

\subsection{Brief history of the dark sector \label{sec:temperatures}}

We shall assume that some unspecified high-energy interactions brought the dark sector plasma in thermal equilibrium with the SM particles at some high temperature, and that these interactions decoupled early, leaving the dark sector at the same temperature as the SM plasma, $\tilde{T} = \tilde{T}_{\rm SM} = \tilde{T}_D$, when all of the SM degrees of freedom where kinetically coupled to the photons.
We also assume that at that common temperature $\tilde{T}$, DM and the dark photon provided the only relativistic degrees of freedom in the dark sector.
Beyond this point, the temperature of the dark photons $\TD$ is in general different from the temperature of ordinary photons $\TSM$. However, $\TD$ (and the DM temperature $\TX$, whenever different) can be computed with respect to $\TSM$ by following the cosmology of the two sectors. This is the goal of this subsection. Determining  $\TD$ and $\TX$ is important for the computation of the DM relic density, the estimation of dark photon cosmological abundance and the associated BBN constraints, and for the late-time DM annihilation and the resulting CMB constraints. We summarise some important events that determine the late-time cosmology in \cref{tab:cosmo}.\footnote{
The two sectors exchange energy via elastic and inelastic scatterings between DM and the SM charged fermions, due to the kinetic mixing of $U(1)_D$ with the hypercharge. The processes $f_{\mathsmaller{\rm SM}}^+ f_{\mathsmaller{\rm SM}}^- \leftrightarrow \bar{X} X$  
dominate the energy transfer and equilibrate the two sectors if 
$\epsilon \gtrsim 3 \times 10^{-6} [\gSM(T\sim \MDM)/10^2]^{1/4} (\MDM/\TeV)^{1/2}(0.03/\alphaD)^{1/2}$~\cite{Chu:2011be}. Since obtaining the observed DM density from freeze-out in the dark sector sets approximately $\alphaD \propto \MDM$ at $\MDM <$~TeV and $\alphaD \propto \MDM^{0.6}$ at $\MDM >$~TeV~\cite{vonHarling:2014kha}, the equilibration condition simplifies roughly to $\epsilon \gtrsim 10^{-6}$, becoming slightly stronger at $\MDM \gtrsim$~TeV. For a large portion of the parameter space considered here (roughly for $\MDM \gtrsim$~few~GeV and $\mVD \lesssim$~GeV), the constraints on $\epsilon$ imposed by direct detection and other experiments preclude the possibility that the two sectors equilibrate via $\epsilon$ (cf.~\cref{fig:eps_mVD}). Moreover, the reach of the indirect probes extends down to $\epsilon \ll 10^{-6}$. 
In order to not curtail the applicability of the indirect detection constraints of \cref{sec:DMID}, and for simplicity, we shall adopt the same assumptions --described in the previous paragraph-- for the decoupling temperature of the two sectors, in the entire parameter space considered.
}
%
\begin{table}[t]
\centering {\small
\renewcommand{\arraystretch}{1.9}
\begin{tabular}{|l|l|} 
\hline
{\bf Temperature} &   {\bf Event}
\\ \hline  \hline 
$T \gtrsim \tilde{T} = \tilde{T}_{\mathsmaller{D}} = \tilde{T}_{\mathsmaller{\rm SM}}$ & 
\parbox[c]{10.5cm}{The dark plasma is in thermal equilibrium with the SM plasma.}
\\ \hline
$\TD^{\rm f.o.} \approx \MDM/30$ & 
\parbox[c]{10.5cm}{The annihilation processes $\bar{X}X \leftrightarrow 2\VD$ freeze-out; the DM density departs from its equilibrium value.}
\\[5pt] \hline
$\TD^{\rm k.d.}$ & 
\parbox[c]{10.5cm}{
\smallskip
The energy transfer via $X \, \VD \leftrightarrow X \, \VD$  becomes inefficient. DM kinetically decouples from the dark photons.}
\\[5pt] \hline
$\TD^{\rm trans} \equiv \max\left[\TD^{\rm k.d.}, \mVD/3\right]$ & 
\parbox[c]{10.5cm}{
The DM temperature transitions from $\TX = \TD \propto 1/a$ to \\ 
$\TX \propto 1/a^2$ scaling. ($a$: scale factor)}
\\[5pt] \hline
$\TD^{\rm decay}$ & The dark photons decay.
\\[5pt] \hline
$\TD^{\rm dom}$ &  
\parbox[c]{10.5cm}{The dark photons would dominate the energy density of the universe, had they not decayed.}
\\[0.2cm] \hline
\end{tabular}}
\caption{\em \small \label{tab:cosmo} Important {\bfseries cosmological mileposts in the evolution of the dark sector}. The subscripts ${}_D$ and ${}_X$ denote the temperatures of the dark photons and DM, respectively.}
\end{table}

\paragraph{Dark photons.} The dark photons are the last particles in the dark sector to become non-relativistic.
So long as the dark sector contains relativistic species, $\TD$ can be easily computed with respect to $\TSM$ by invoking the conservation of the comoving entropy in each sector separately. As is standard, we find
\beq
r \equiv \frac{\TD}{\TSM}
= \left[\frac{\gSM(\TSM)}{\gD(\TD)}\right]^{1/3} \, 
\left(\frac{\gDtilde}{\gSMtilde} \right)^{1/3} \,,
\label{eq:r}
\eeq
where $\gSM$ and $\gD$ stand for the relativistic degrees of freedom in the SM sector and the dark sector respectively, and $\tilde{g}$ refers to their values at the common temperature $\tilde{T}$. 
After the dark photons become non-relativistic, at $\TD \approx \mVD/3$, their momentum redshifts as $p \propto 1/a$, where $a$ is the scale factor of the universe, and their distribution resembles a thermal one with temperature that now scales as $\TD \propto p^2 \propto 1/a^2$. Since no interactions that can change the dark photon number are in  equilibrium, the dark photons develop a non-zero chemical potential. Both their comoving number and entropy are conserved.\footnote{\label{foot:NRthermo}
We recall that for a non-relativistic species of mass $m$, with non-zero chemical potential $\mu$, at temperature $T$, the number and entropy densities are $n = g [m T/(2\pi)]^{3/2} \, \exp[(\mu - m) /T]$ and $s = n (m-\mu)/T$. The scaling $T\propto 1/a^2$ and $(\mu -m) \propto T$ ensures the conservation of the comoving number and entropy: $n,s \propto 1/a^3$.}

\paragraph{Dark matter.} Early on, DM is in kinetic and chemical equilibrium with the dark photons, via elastic and inelastic scatterings, $X\,\VD \leftrightarrow X\,\VD$ and $\bar{X}\,X \leftrightarrow \VD \, \VD$. The DM density follows (approximately) an equilibrium distribution, with temperature $\TX = \TD$. At temperature $\TX^{\rm f.o.} \approx \MDM/30$, the inelastic scatterings fail to keep DM in chemical equilibrium with the dark photons, and the DM density freezes-out. (However, the full chemical decoupling of $X$ and $\VD$ may occur at a somewhat later time, cf.~\cref{sec:relic}.) After freeze-out, DM remains in kinetic equilibrium with the dark photons, via elastic scatterings. It continues to have an equilibrium density with $\TX=\TD$, albeit it develops  a non-zero chemical potential. This remains true, up until the time of kinetic decoupling of DM with the dark photons, at a temperature $\TD^{\rm k.d.}$ that we shall estimate below. After the kinetic decoupling, the momentum of the DM particles redshifts as $p \propto 1/a$. Their population continues to resemble an equilibrium distribution, with a chemical potential that evolves to reflect the conservation of the particle number, and a temperature that scales as $\TX \propto p^2 \propto 1/a^2$ (and departs from the temperature of the dark photons, if the latter remain relativistic at that time).

Putting the above considerations together, we conclude that the DM temperature is
\beq
\TX \approx \left\{
\begin{aligned}
&r(\TSM) \ \TSM  \,, 
&\quad \text{for} \quad \TD > \TD^{\rm trans} \,,
\\
& \frac{(1+z)^2}{(1+z_{\rm trans})^2} \ \TD^{\rm trans} \,, 
&\quad \text{for} \quad \TD < \TD^{\rm trans} \,,
\end{aligned}
\right.
\label{eq:TX}
\eeq
where the temperature $\TD^{\rm trans}$ marks the transition from the $\TX \propto 1/a$ to the  $\TX \propto 1/a^2$ scaling, that occurs either when DM decouples kinetically from the dark photons, or when the dark photons become themselves non-relativistic, whichever happens first,
\beq 
\TD^{\rm trans} \equiv \max[\TD^{\rm k.d.}, \, \mVD/3] \,.
\label{eq:TDtrans_def}
\eeq
In \cref{eq:TX}, $z$ denotes the redshift. 
We determine $z_{\rm trans}$ from the conservation of entropy in the SM sector, $\gSM \TSM^3 (1+z)^{-3} = g_0 T_0^3$, where the index ``0" refers to today's values, as typical. Using also \cref{eq:r}, we find
$1+z_{\rm trans} = (\TD^{\rm trans} / T_0) \times
(\gD^{\rm trans} \, \gSMtilde)^{1/3} 
/ (g_0 \gDtilde)^{1/3}$.
We shall take $\gSMtilde =106.75$ to account for all the SM degrees of freedom, $\gDtilde = 7$ to account for the massive dark photon and the DM degrees of freedom, and $\gD^{\rm trans}= 3$. Then, 
\beq 1+z_{\rm trans} \approx 2.5\,(\TD^{\rm trans} / T_0) \,. \label{eq:ztrans} \eeq 
We shall use \cref{eq:TX,eq:ztrans} in \cref{sec:CMB}, to obtain constraints from the DM annihilation around CMB. We determine $\TD^{\rm trans}$ next.

\paragraph{Kinetic decoupling and transition temperature.} 
The DM and the dark photons are kept in kinetic equilibrium via the elastic scatterings $X\, \VD \leftrightarrow X\,\VD$. Assuming that the dark photons are relativistic, this process is described by the Thomson cross-section $\sigma_T = 8\pi \alphaD^2/(3\MDM^2)$. The energy transfer rate is 
$d\rho/dt \simeq (\delta E_1) \, n_{\mathsmaller{X}} n_\mathsmaller{\VD} \sigma_T $, 
where $n_{\mathsmaller{X}}$ is the density of the DM particles, $n_{\mathsmaller{\VD}} = [\zeta(3)/\pi^2] \gVD \TD^3$ is the number density of the dark photons, and  $\delta E_1\approx \TD^2 /(2\MDM)$ is the average energy transfer per collision (with $\delta p_1 \sim \TD$ being the average momentum transfer). The kinetic decoupling occurs when 
$\rho_{\mathsmaller{X}, \rm kin}^{-1} (d\rho/dt) < H$, where 
$\rho_{\mathsmaller{X}, \rm kin} \simeq (3/2) \TD n_\mathsmaller{X}$ is the DM kinetic energy density, and $H$ is the Hubble parameter. Putting everything together, we find that the kinetic decoupling occurs at 
$\TD^{\rm k.d.} \approx 2(\MDM^3 /M_{\rm Pl})^{1/2} /(\alphaD \, r_{\rm k.d.})$. Our calculation breaks down if the estimated temperature is  $\TD^{\rm k.d.} \lesssim \mVD/3$. Since we are interested in $\TD^{\rm trans}$ rather than $\TD^{\rm k.d.}$, we need not repeat the calculation for non-relativistic dark photons. We conclude that
\beq
\TD^{\rm trans} \approx 
\max\left[
\frac{2\MDM^{3/2}}{r_{\rm trans} \, \alphaD \, M_{\rm Pl}^{1/2}}, \ \frac{\mVD}{3}
\right] \,,
\label{eq:TDtrans}
\eeq
where we estimate $r_{\rm trans} \sim 0.6 - 1$ using \cref{eq:r}. (For a more precise treatment of the kinetic equilibrium and decoupling, see ref.~\cite{vandenAarssen:2012ag}.)

\paragraph{Dark photon mass generation.} 
In the discussion above, we have not been concerned about the generation of the dark photon mass. If $\mVD$ is generated via the St\"uckelberg mechanism, then the dark photon is massive throughout the cosmological history. However, if the dark photon obtains its mass via a dark Higgs mechanism, then it becomes massive only after a $U(1)_D$ - breaking phase transition takes place. We estimate the temperature of the latter to be 
$\TD^{\rm p.t.} \sim v_{\mathsmaller{\rm D}} = \mVD / \sqrt{8\pi\alphaD}$, where $v_\mathsmaller{D}$ is the vacuum expectation value of dark Higgs. We see that for the entire range of $\alphaD$ considered, $\TD^{\rm p.t.} \gtrsim \mVD /3$, i.e.~the dark photons are at least quasi-relativistic after the phase transition, and our considerations are left largely unaffected. 

Of course, a dark Higgs would imply one extra degree of freedom in the dark sector, in particular $\gDtilde = 8$ and $\gD^{\rm f.o.} =$~3~or~4, depending on the mass of the dark Higgs. This would alter $\TD$ at the time of DM freeze-out, and hence the required annihilation cross-section, only by $\sim 5\%$. Moreover, it would not change significantly the expansion rate of the universe, which is driven mostly by the SM degrees of freedom at all times. Our results, produced under the assumptions described above, are thus valid for both the St\"uckelberg and the Higgs mechanisms.

\subsection{Relic abundance \label{sec:relic}}

\begin{figure}[!t]
\begin{center}
\includegraphics[height= 0.48 \textwidth]{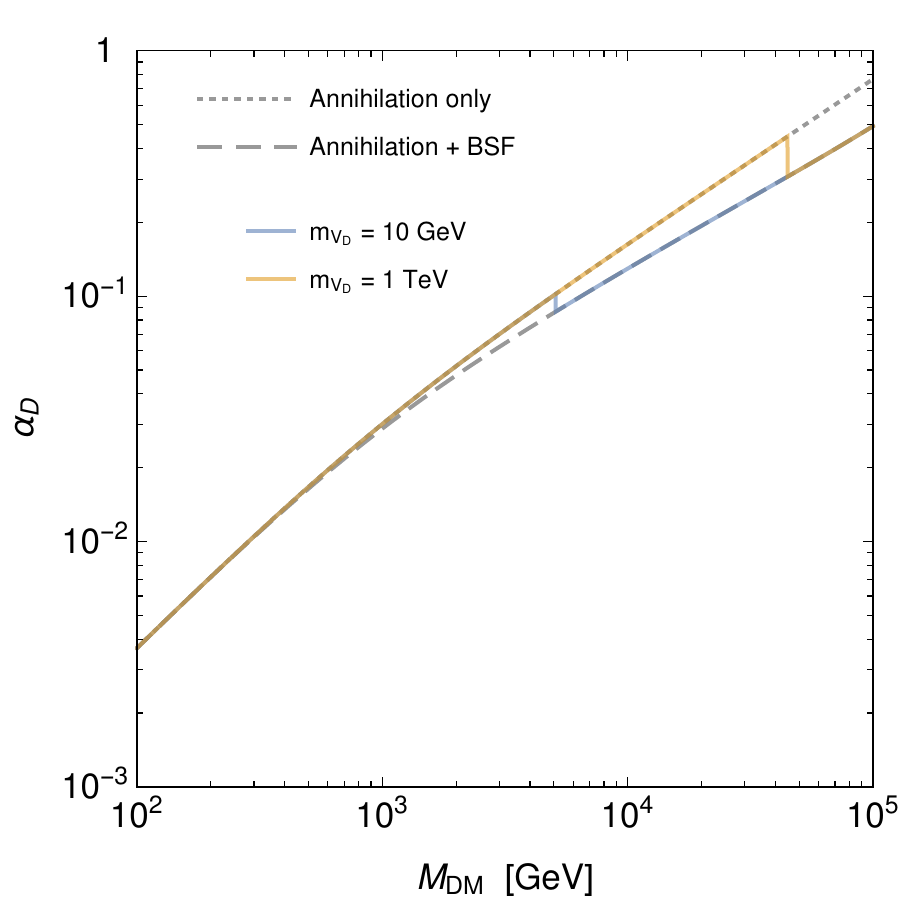}~~
\includegraphics[height= 0.48 \textwidth]{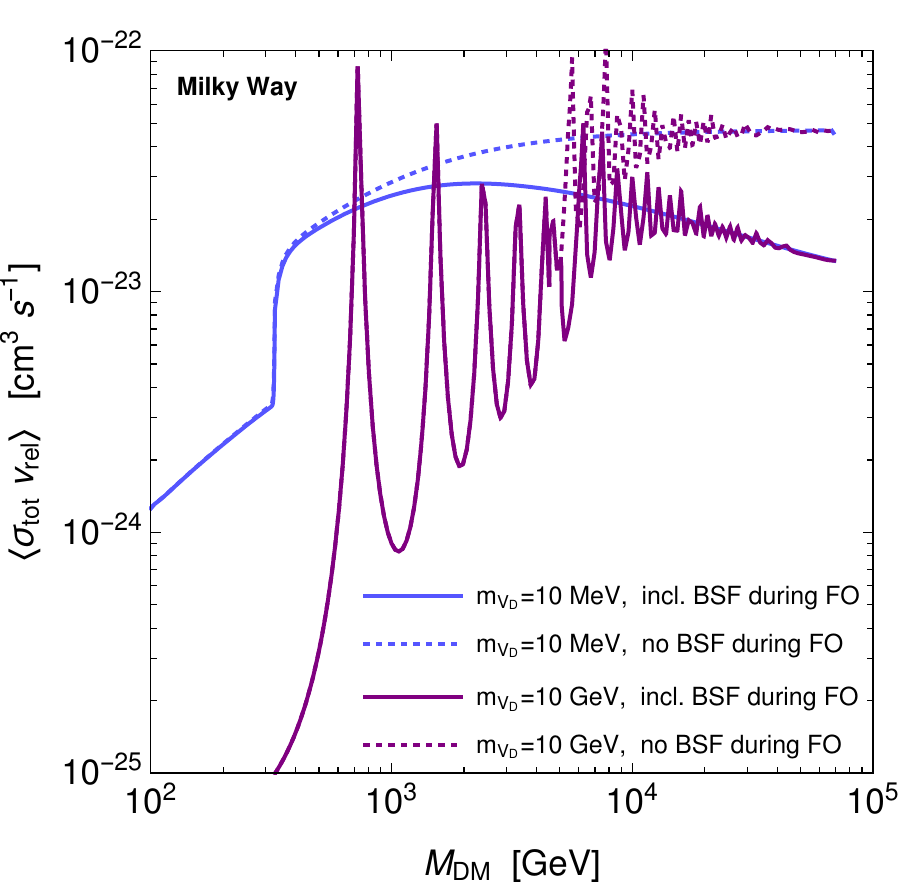}
\caption[]{\em \small \label{fig:alpharelic} 
Left: {\bfseries The dark fine structure constant} $\alphaD$ is determined by the thermal freeze-out of DM in the dark sector. 
The dotted gray line takes into account annihilation only, while the 
dashed gray line incorporates both annihilation and bound-state formation. 
The overprediction of $\alphaD$ if BSF is neglected is $(10-55)\%$ for DM masses $(3-100)$~TeV. 
Both calculations have been performed in the Coulomb limit, 
which is a good approximation during freeze-out, except for 
the phase-space suppression due to a dark photon emission in 
the case of BSF. Thus, for non-zero dark photon masses, we interpolate between the two calculations near the threshold where BSF becomes kinematically possible, as shown by the coloured solid lines. 
Right: {\bfseries Implications of BSF during freeze-out, for the DM indirect detection signals today.} We show the total cross-section (annihilation plus BSF), averaged over Milky Way velocities. The dotted lines correspond to $\alphaD$ determined by considering only annihilation during freeze-out. The solid lines correspond to $\alphaD$ determined via the full calculation that includes both annihilation and BSF. The effect is two-fold: A lower $\alphaD$ implies lower overall rates for the expected signals today, and shifts the resonances to larger values of the DM mass.}
\end{center}
\end{figure}

We determine the dark fine structure constant $\alphaD$ by requiring that the observed DM abundance is attained after the thermal freeze-out of DM from the dark photon plasma. For large $\mVD$, the interaction between DM and dark photons is contact-type, and $\sigma_{\rm ann} v_{\rm rel} \simeq \sigma_0$ is velocity independent. As is standard, the predicted value for $\sigma_0$ is then essentially independent of the DM mass,~\footnote{A step-like variation at $\MDM \lesssim 10$~GeV arises due to the decoupling of the QCD degrees of freedom during freeze out (see e.g.~\cite{Steigman:2012nb}).} 
thus setting $\alphaD \propto \MDM$ (as it trivially follows from the definition of $\sigma_0$ in sec.~\ref{sec:cross-sections}).
However, in the parameter space where the condition~\eqref{eq:longrange} in satisfied,  the annihilation processes are Sommerfeld enhanced. This affects significantly the DM relic density for $\MDM \gtrsim$~TeV~\cite{Hisano:2006nn}. More recently, it was shown that, in the same mass range, the formation and decay of particle-antiparticle bound states depletes significantly the DM abundance~\cite{vonHarling:2014kha}, thereby reducing the predicted $\alphaD$ further.

We follow the analysis of ref.~\cite{vonHarling:2014kha}, to determine the dark fine structure constant $\alphaD$. We perform our calculations in the Coulomb regime, using eqs.~\eqref{eqs:Sfactors_Coulomb} to determine the annihilation and BSF cross-sections, and we verify \emph{a posteriori} that this is a satisfactory approximation (cf.~\cref{app:RelicDensity_Coulomb}). We compute $\alphaD$ in two different cases: first, considering the DM direct annihilation only, and then considering both direct annihilation and bound-state formation and decay. The latter requires solving a set of Boltzmann equations that capture the interplay of bound-state formation, ionisation and decay processes~\cite{vonHarling:2014kha}. 
However, for a massive dark photon, BSF is not always kinematically allowed. 
To account for this kinematic cutoff and the phase-space suppression, in our calculations of the various DM interaction rates throughout our work, we adopt the former computation of $\alphaD$ when  $\pss^{1/2} (3-\pss)/2 < 0.5$, and the latter when $\pss^{1/2} (3-\pss)/2 > 0.5$, where $\pss$ is defined in \cref{eq:pss} [see also~\cref{eq:sigma_BSF}]. 
\Cref{fig:alpharelic} (left panel) shows $\alphaD$ as a function of the DM mass.\footnote{
While in \cref{fig:alpharelic} we present our results only for $\MDM \gtrsim 100$~GeV, we carry out our computations for the entire mass range considered in this work, 1~GeV~$\lesssim \MDM \lesssim$~100~TeV.  We note that we improve with respect to the computation of ref.~\cite{vonHarling:2014kha}, by taking into account the variation of the dark-to-ordinary plasma temperature ratio, due to the decoupling of the SM and DM degrees of freedom, according to \cref{eq:r}. See ref.~\cite{Baldes:2017gzw} for the Boltzmann equations in terms of the dark sector temperature.}

Note that $\alphaD$ may be lower than estimated here, if in the early universe the dark sector was at a significantly lower temperature than the SM plasma. This would imply overall more relaxed direct and indirect detection constraints, as well as weaker DM self-interactions. However, a significantly colder dark sector requires either a large number of yet unknown particles with sizeable couplings to the SM, whose cosmological decoupling increased the entropy of the SM plasma after its decoupling from the dark sector, or appropriate initial conditions set by inflation and absence of any interactions that would equilibrate the two sectors (see e.g.~\cite{Adshead:2016xxj}). We do not explore this possibility further in this paper.

\paragraph{Phenomenological implications of the BSF effect on the DM density.}

Neglecting BSF in the determination of the DM relic density overpredicts $\alphaD$ by about $(15-55)\%$ in the DM mass range $(5-100)$~TeV. This overestimates the direct detection constraints on the DM-nucleon cross-section, and consequently on $\epsilon^2$, by the same amount. The effect on the DM annihilation and BSF cross-sections relevant for indirect searches and CMB constraints, is greater. In the Coulomb regime, which pertains to much of the parameter space of the model for indirect searches in the  Milky Way and the Dwarfs (cf.~\cref{fig:PhaseSpace}), as well as at non-resonant points away from the Coulomb regime, the inelastic cross-sections scale as 
$\sigma_{\rm ann, \, BSF} \propto \alphaD^3$. Then, the overestimation of the expected indirect detection signals ranges from 50\% up to a factor of $\sim 4$, in the same DM mass interval. The discrepancy is also stark at resonant points, which appear at discrete values of $\alphaD \MDM/(2\mVD)$, as described in \cref{sec:cross-sections}; the overestimation of $\alphaD$ implies that the resonances would be  expected to occur at lower $\MDM$ values. These effects are depicted in \cref{fig:alpharelic} (right panel).

\subsection{BBN constraints \label{sec:BBN}}

\begin{figure}[!t]
\begin{center}
\includegraphics[width= 0.48 \textwidth]{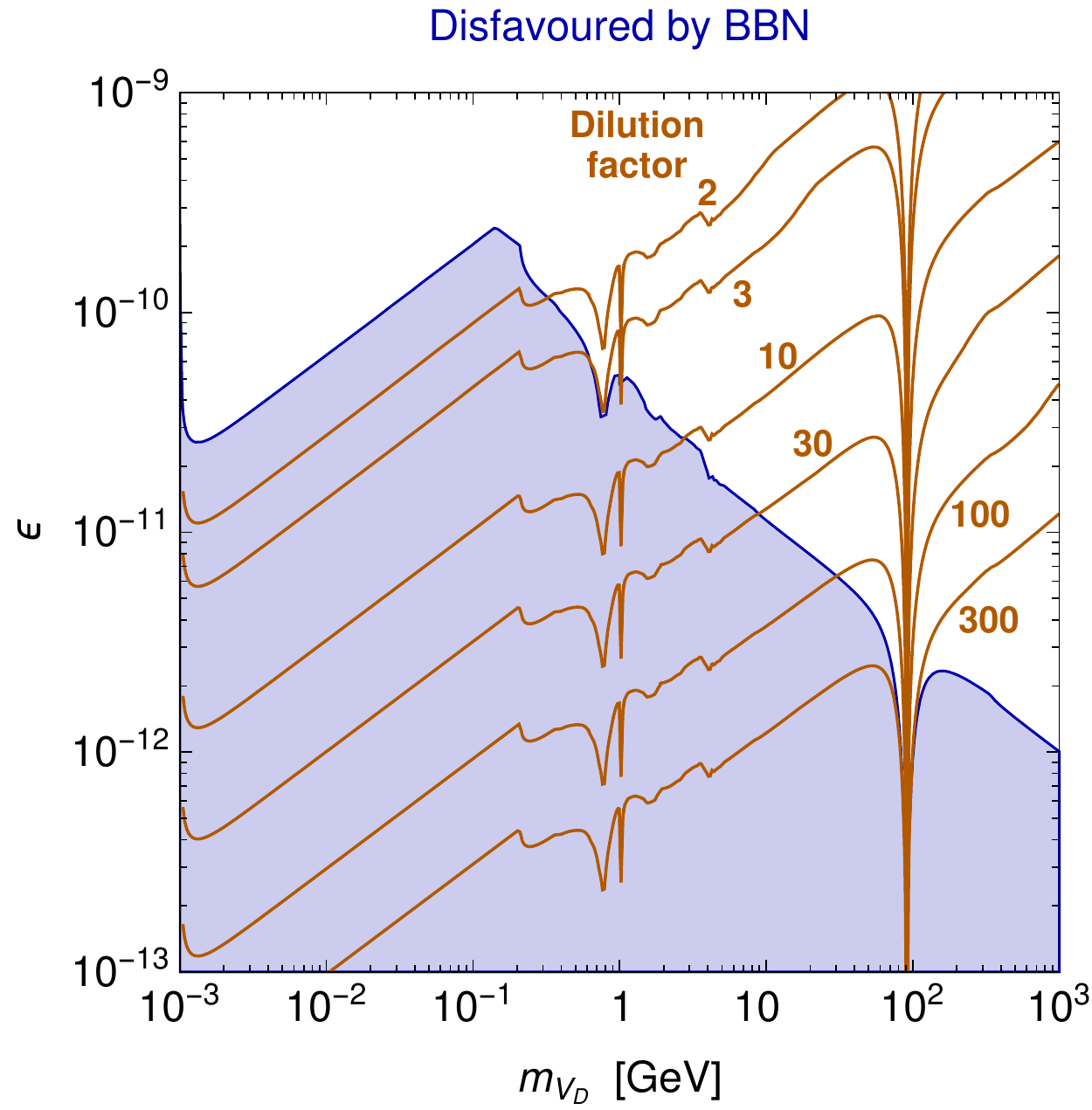}~~
\caption[]{\em \small \label{fig:cosmology_bounds} {\bfseries Regions disfavoured by BBN}. 
Outside the shaded area, the dark photons either decay safely before BBN ($\tau_{\VD}^{} \lesssim 0.03$s above border to the right), or their cosmological abundance is not significant (above border to the left). 
The disfavoured area is a rough estimate only (see text for discussion). 
For large $\mVD$ and small $\epsilon$, the decay of the dark photons injects significant entropy in the universe and dilutes DM. The solid orange lines indicate the estimated dilution factor. In this regime, $\alphaD$ should be significantly lower than estimated, in order for the dark fermions to account for the observed DM abundance. This would relax the CMB and indirect detection constraints.}
\end{center}
\end{figure}

The cosmological abundance of dark photons freezes-out essentially at the time of DM freeze-out, when the interaction that can alter their number become inefficient. After the dark photons become non-relativistic, they may dominate the energy density of the universe, and cause cosmological problems. In particular, they may inject a large amount of entropy if they decay around or after BBN, and/or shift the time of matter-radiation equality. Here, we estimate the constraints implied by these considerations.

Assuming they were relativistic at the time of DM freeze-out ($\mVD/3 \lesssim \TD^{\rm f.o.} \approx \MDM/30$), the ratio of the dark photon number density to SM entropy density was
\beq
\fVD =\frac{\nVD}{\sSM} \approx 
\frac{45\zeta(3)}{2\pi^4} \frac{\gVD \, r_{\rm f.o.}^3 }{\gSM^{\rm f.o.}} = 
\frac{45\zeta(3)}{2\pi^4} \frac{\gDtilde }{\gSMtilde}
\,, \label{eq:fVD}
\eeq
where $r_{\rm f.o.}$ is the dark-to-ordinary temperature ratio at the time of DM freeze-out, computed according to \cref{eq:r}, and we used $\gD^{\rm f.o.}=\gVD$. From \cref{eq:fVD}, we estimate $\fVD \sim 0.02$. 
In the absence of decay, the dark photons are set to exceed a fraction $f$ of the energy density of the universe when $\mVD \fVD \sSM \gtrsim f \, (\pi^2/30) \gSM \TSM^4$, or
\beq
\TSM^{\rm dom} \approx (4 \mVD \fVD) /(3 f) \,.
\label{eq:Tdom}
\eeq

An early time of matter domination induced by dark photons, is not in conflict with late-time cosmology, provided that radiation domination is re-established before BBN. We shall thus require that the dark photons either decay before BBN, or that their energy density does not exceed a critical fraction $f$ of the energy density of the universe prior to their decay, 
\beq
\tauVD \lesssim \max [\tauBBN, H(\TSM^{\rm dom})^{-1}]
\label{eq:BBNconstraint}
\eeq
where $\tauVD$ is the dark photon lifetime computed according to \cref{sec:BRs}, and $H(\TSM^{\rm dom})$ is the Hubble parameter at the temperature $\TSM^{\rm dom}$ estimated from \cref{eq:Tdom}.

The parameter space disfavoured by the condition \eqref{eq:BBNconstraint} is illustrated in \cref{fig:cosmology_bounds}. We have taken $f \sim 0.5$ and $\tauBBN \simeq 0.03$~s. This choice of $\tauBBN$ is motivated by the known BBN constraints on dark photons; our disfavoured region encompasses all the parameter space that was excluded by a proper BBN analysis in ref.~\cite{Berger:2016vxi}, under minimal assumptions about the dark photon cosmological density. In the our case, the dark photon primordial density can be significantly larger. Nevertheless, the effect of dark photon decay on BBN is intricate, and depends not only on the dark photon energy density and decay rate, but also on their mass, which determines their decay channels, and possibly on other details~\cite{Berger:2016vxi}. Our constraints are thus a rough estimate only; it is possible that some parts of the parameter space close to the border of the disfavoured region are in fact allowed. The derivation of more precise BBN constraints is beyond the scope of our work.

The above constraints are evaded if the dark photons are sufficiently light. In this case, their energy density redshifts as radiation sufficiently long and does not become significant by the time of matter-radiation equality. Then, the dark photons eventually make up only a subdominant component of DM (or radiation, if extremely light or massless). Requiring that $\TSM^{\rm dom} \lesssim T_{\rm eq} \simeq 0.8~\eV$, we find that this occus for $\mVD \lesssim 15~\eV$. This is the equivalent of the Gershtein-Zel'dovich/Cowsik-McClelland bound~\cite{Gershtein:1966gg,Cowsik:1972gh}, adapted to our setup.

\subsection{Entropy production from dark photon decay \label{sec:entropyprod}}

If the dark photons decay before BBN, but after they constitute a sizeable fraction of the energy density of the universe, i.e.~if $H(\TSM^{\rm dom})^{-1} \lesssim \tauVD \lesssim \tauBBN$, their decay generates significant entropy, which dilutes any decoupled relic, including DM. We may estimate the dilution factor using the simultaneous decay approximation, and invoking the conservation of energy before and after the decay (see e.g.~ref.~\cite{Scherrer:1984fd}),
\begin{subequations}
\label{eqs:rhos}
\begin{align}
\rho_{\rm before} 
&= (\pi^2 / 30) \ \gSM \ {\TSM^{\rm before}}^{\,4}
+ \mVD \fVD \ (2\pi^2/45) \ \gSM \ {\TSM^{\rm before}}^{\,3} \,,
\label{eq:rho_before}
\\
\rho_{\rm after} 
&= (\pi^2 / 30) \ \gSM \ {\TSM^{\rm after}}^{\,4} \,,
\label{eq:rho_after}
\\
\rho_{\rm before} = \rho_{\rm after} 
&= 3M_{\rm Pl}^2 / (8\pi \tauVD^2)  \,,
\label{eq:rho_before=after}
\end{align}
\end{subequations}
where $\fVD$ has been estimated in \cref{eq:fVD}. 
The entropy densities are
\begin{subequations}
\label{eqs:s_BeforeAfterDecay}
\begin{align}
s_{\rm before} &= (2\pi^2 / 45) \ \gSM
\left(1 + \gVD r_{\rm f.o.}^3 / \gSM^{\rm f.o.} \right)
\ {\TSM^{\rm before}}^{\,3} \,,
\label{eq:s_before}
\\
s_{\rm after}  &= (2\pi^2 / 45) \ \gSM \ {\TSM^{\rm after}}^{\,3} \,,
\label{eq:s_after}
\end{align}
\end{subequations}
where in \cref{eq:s_before} we have accounted both for the SM sector and the dark photon entropy densities (see \cref{foot:NRthermo}). We determine the temperatures $\TSM^{\rm before}$ and $\TSM^{\rm after}$ from eqs.~\eqref{eqs:rhos}, and estimate the dilution factor to be
\beq
\frac{s_{\rm after}}{s_{\rm before}} \approx 
\frac
{1+ 1.72 \fVD \, \gSM^{1/4} \mVD \sqrt{\tauVD/M_{\mathsmaller{\rm Pl}}}}
{1 + 3.6\fVD } \,.
\label{eq:S_ratio}
\eeq

\Cref{fig:cosmology_bounds} shows that the dilution can be very significant for $\mVD \gtrsim 1$~GeV and small $\epsilon$.\footnote{
Note that because in reality, the entropy injection from the dark photon decay does not reheat the universe, but it rather makes it cool more slowly~\cite{Scherrer:1984fd}, the shaded region with large dilution factor in \cref{fig:cosmology_bounds} cannot be un-excluded by claiming that the temperature is reset to a pre-BBN value.}
In this event, the $X$ fermions can account for the observed DM density only if they freeze-out with a significantly larger abundance. This requires a smaller $\alphaD$ than that estimated in \cref{sec:relic} in the absence of any excessive entropy production, which in turn implies overall smaller annihilation and BSF cross-sections at late times, thereby relaxing the CMB and indirect detection constraints. The possibility of a significant late entropy injection in this model has already been pointed out in ref.~\cite{Berlin:2016vnh}, which however focused on a different region of the parameter space. We leave a more detailed study of the implications of this effect for future work.

\section{Constraints from indirect DM searches and the CMB}
\label{sec:DMID}

In this section we move to the indirect detection probes of the model outlined above. In the commonly assumed sense, indirect detection strategies aim at revealing excesses and features in cosmic ray fluxes collected at Earth, which could be ascribed to DM annihilations. The searches focus in particular on regions where DM is most dense or where the astrophysical backgrounds are particularly reduced. Here we concentrate on three indirect detection probes: gamma ray searches with the {\sc Fermi} satellite in the Milky Way galactic halo and in dwarf spheroidal galaxies (\cref{sec:gammas}) and antiproton searches with the {\sc Ams-02} experiment (\cref{sec:antiproton}). 

In a broader sense, the CMB is also a powerful indirect probe, since it is sensitive to DM annihilations into SM particles at the time of recombination. Such annihilations inject energy in the plasma, causing ionisation and heating of the medium, production of low energy photons, etc. Ionisation is particularly important: the increased amount of free electrons affects the CMB anisotropies, and makes them potentially inconsistent with observations (see refs.~\cite{Padmanabhan:2005es, Galli:2009zc, Slatyer:2009yq, Huetsi:2009ex, Cirelli:2009bb} for the first DM studies using {\sc Wmap} data).

The compilation of the constraints that we derive in this section is presented in fig.~\ref{fig:allbounds}.
\medskip

As mentioned in the introduction, DM bound states will annihilate into 2 or 3 dark mediators $\VD$, depending on the spin state. In turn, $\VD$ decays into pairs of SM particles such as $e^+e^-, \mu^+\mu^-$, light quarks and heavy quarks, according to the BRs discussed in detail in \cref{sec:BRs}. Such SM states will then shower and hadronise, when applicable, producing fluxes of stable particles ($\gamma$-rays, $e^\pm$ and $\bar p$) which can be compared meaningfully with the observations in gamma-rays and antiprotons. We take these spectra from the detailed work in~\cite{Elor:2015bho}, where they are provided for a variety of channels and for a large range of masses. Their results are based on previous {\sc Pppc4dmid} and {\sc Pythia} computations.\footnote{The numerical results are also linked from the \href{http://www.marcocirelli.net/PPPC4DMID.html}{{\sc Pppc4dmid} website}.} A few technical comments are however in order. i)~While the provided spectra are computed for the case of scalar mediators, they can also be adopted for the case of vector mediators in which we are interested, as discussed in particular  in~\cite{Elor:2015tva}. ii)~The spectra for the decay of spin-singlet states in 2 $\VD$ are given explicitly, while those for spin-triplet states in 3 $\VD$ have to be approximately derived. As discussed in detail in~\cite{Elor:2015bho}, the direct decay into 2$n$ SM states is roughly equivalent to the decay via an $n$-step cascade, at least for $n=1,2$. We therefore determine the spectra for our 3-step cascade by averaging the $n=1$ and $n=2$ spectra, and verify that, for our purposes, this is effectively the same of taking the $n=2$ spectra. We further check, in specific cases, that our averaged spectra are in very good agreement with those obtained by the analytical form given in ref.~\cite{Elor:2015bho} (eq.~(A4) therein). iii)~The spectra from~\cite{Elor:2015bho} are parametrized in terms of the quantity $2 \, m_{\rm SM}/\mVD$ (with $m_{\rm SM}$ the mass of the outgoing SM particles, if non vanishing), within a limited range: when needed (e.g. for large $\mVD$) we will linearly extrapolate the spectra and check the consistency of the procedure.  iv)~The spectra for light quarks are not provided in~\cite{Elor:2015bho}: we use the ones for gluon pairs, given their close similarity. Such spectra are given for $\mVD \ge 10$ GeV so we extrapolate down to 5 GeV, below which we adopt the refined treatment discussed in \cref{sec:BRs}.

For the CMB, the impact on the medium depends on the fraction $f$ of the injected power (the products of DM annihilation) that is actually deposited in the environment and thus matters for reionisation. 
This `efficiency factor' $f$ is, in general, a function of the DM annihilation channel, the injection energy, and the injection redshift $z$. Reference~\cite{Slatyer:2015jla} demonstrated however that it is possible to account for the redshift dependence with a constant effective function $f_{\rm eff}$, and allowed to compute such a function for a wide set of injection energies (\textit{i.e.} DM masses) and arbitrary annihilation channels. We will use this quantity in the following.

\subsection{Gamma-ray searches}
\label{sec:gammas}

\paragraph{Dwarf spheroidal galaxies (dSphs).} These objects are among the best targets to look for DM signals with gamma-ray observations. So far, no  evidence for an excess of photons over the background has been found. This allows to set stringent constraints on the DM annihilation cross-section. 
For this purpose, we follow closely the analysis performed by the {\sc Fermi} collaboration, which is based on the observation of 15 dwarf galaxies in the energy range  500 MeV-500 GeV~\cite{Ackermann:2015zua}. The statistical analysis is performed as follow.

For each target, the likelihood $\mathcal L_i(\boldsymbol{\mu}, J_i | \mathcal D_i)$ depends on the gamma-ray data, $\mathcal D_i$, the parameters of the model $\boldsymbol{\mu}$ (i.e. the particular point in the ($\MDM, \mVD$) parameter space under consideration) and the so-called $J$-factors $J_i$ .
The latter quantities correspond to the integral of the square of the DM density profile in the window of the observation. These parameters, specific for each dwarf galaxy, are needed to determine the DM fluxes. 
For each dwarf galaxy, {\sc Fermi} provide the likelihood in different energy bins as a function of the integrated signal flux. Thus, for a given point $\boldsymbol{\mu}$ in the parameter space and choice of $J_i$, we compute the DM flux in each energy bin and determine the corresponding likelihood. 
Then, we simply multiply these likelihoods for all the energy bins in the analysis. This gives the likelihood $\tilde{\mathcal L_i}(\boldsymbol{\mu}, J_i | \mathcal D_i)$. The total likelihood $\mathcal L_i(\boldsymbol{\mu}, J_i | \mathcal D_i)$ is obtained  multiplying by an additional contribution:

$$\mathcal L_i(\boldsymbol{\mu}, J_i | \mathcal D_i)=\tilde{\mathcal L_i}(\boldsymbol{\mu}, J_i | \mathcal D_i)\times \frac{1}{\ln(10) J_{{\rm obs},i} \sqrt{2 \pi}\sigma_i} e^{-(\log_{10}(J_i)-\log_{10}(J_{{\rm obs},i}))^2/2\sigma_i^2}$$

\noindent This method allows to take into account the uncertainty on the determination of the $J$-factors. We take the measured values $J_{{\rm obs},i}$ and uncertainties $\sigma_i$ from~\cite{Ackermann:2015zua}. 
Finally, we combine the likelihoods of all the 15 dwarf galaxies and treat the $J$-factors $J_i$ as nuisance parameters. We can then determine the regions of the parameter space excluded at 95\% CL performing a test-statistic, comparing the likelihood with and without the DM signal.

We recall that the annihilation and BSF cross-sections are computed by averaging over the speed distribution of DM inside the dwarf galaxies as described in sec.~\ref{sec:cross-sections}. 

\smallskip

\begin{figure}[!t]
\begin{center}
\includegraphics[width= 0.48 \textwidth]{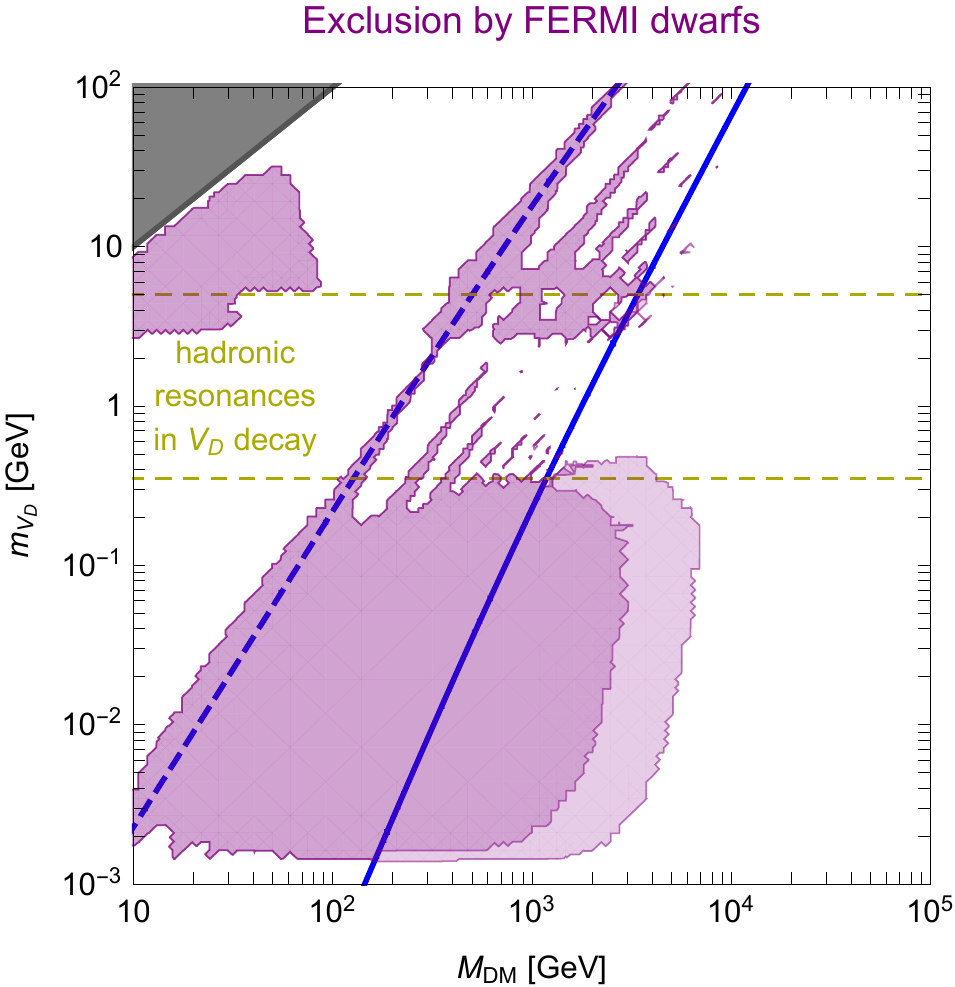}
\includegraphics[width= 0.48 \textwidth]{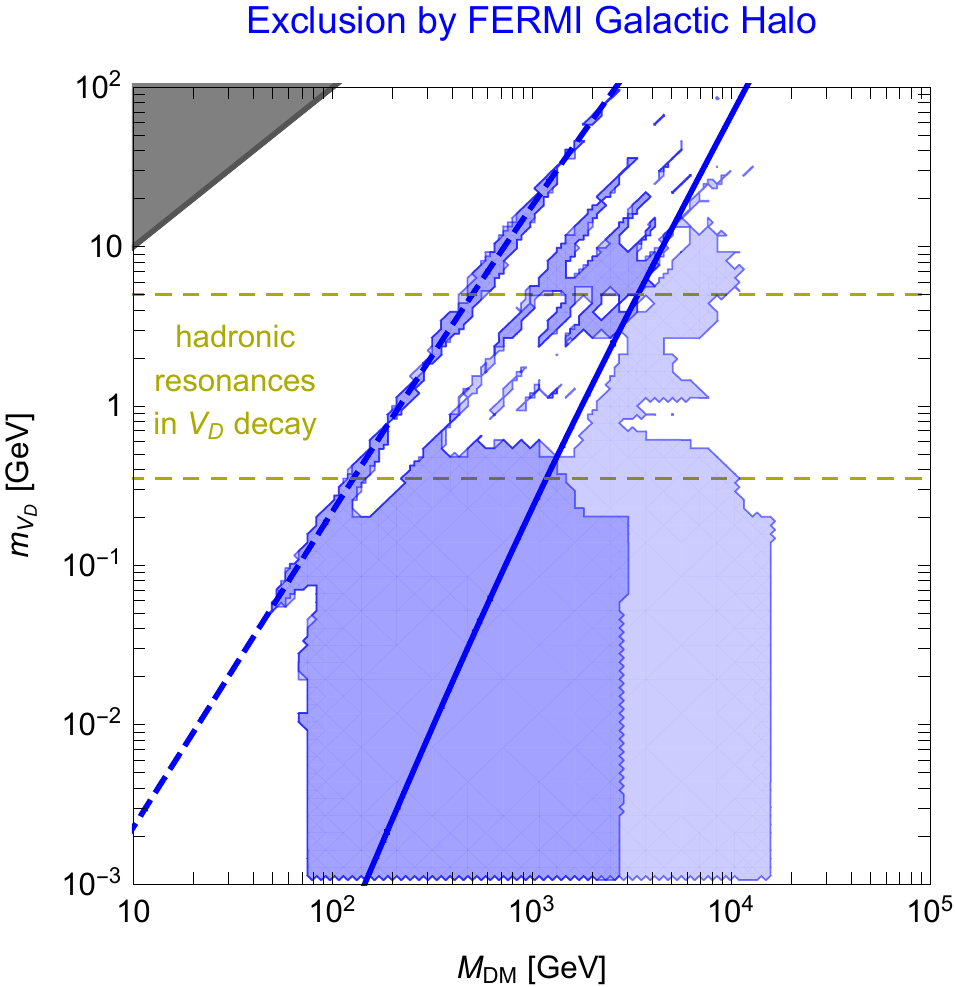}
\caption{\em \small \label{fig:gammas} {\bfseries Gamma-ray constraints} from dwarf galaxies (left) and from  the Galactic Halo (right) in the plane $\MDM/\mVD$. If neglecting the phenomenon of bound state formation, only the darker shaded region is probed. The inclusion of bound states extends the area.}
\end{center}
\end{figure}

Our results are  shown in \cref{fig:gammas} left. The presence of resonances of the annihilation cross-section are clearly visible. At low DM masses ($\MDM\lesssim 10$ GeV), only values of $\mVD \gtrsim$ 3 GeV are excluded. This is because  lighter  dark photons decay mostly/only into leptonic channels, producing less pronounced gamma-ray spectra than those from hadronic channels. At larger DM masses this effect is compensated by a larger annihilation cross-section and by the fact that the photon spectrum shifts at higher energies.The consequence of the formation of bound states can be isolated comparing the lighter and darker regions of \cref{fig:gammas}, left. The inclusion of this effect rules out an additional region of the parameter space ($2\lesssim \MDM\lesssim 5$~TeV).

\bigskip

\paragraph{Milky Way galactic halo (GH).} We derive another set of constraints from {\sc Fermi} observations of the diffuse gamma-ray emission from our  Galaxy. The strategy that we adopt is based on the analysis in~\cite{Cirelli:2015bda}. We consider two mid-latitude regions of interest (RoI 12 and 24 in~\cite{Cirelli:2015bda}) defined by $|\ell|<80^{\circ}$ and $5^{\circ}<b<15^{\circ}$ ($-15^{\circ}<b<-5^{\circ}$) for the region 12 (24), with $\ell$ and $b$ the galactic longitude and latitude. We model the astrophysical emission inside these RoIs as a superposition of different templates. We include 
i) the emission from the interactions of the cosmic-rays with the interstellar radiation field and the gas, 
ii) a template for point sources, 
iii) a template for the so-called `{\sc Fermi} bubbles' and 
iv) the isotropic gamma-ray background.  For the DM signal we include both the primary gamma-rays produced by the decays of the dark photon and the Inverse Compton secondary emission (for this purpose we use the tool in~\cite{Cirelli:2010xx,Buch:2015iya}).
We derive constraints on the model comparing the signal and the background fluxes in the RoIs with {\sc Fermi} observations. We exclude the points with $\Delta \chi^2>9$ with respect to the background-only hypothesis.   We refer to~\cite{Cirelli:2015bda} for further details.
As explained in~\cite{Cirelli:2015bda}, we limit our analysis to energies above $\sim1.5$ GeV, since only for these energies our background model provides a reliable description of the diffuse emission. We average the annihilation and BSF cross sections considering the Maxwellian speed distribution described in sec.~\ref{sec:cross-sections}. 

\smallskip

The results are shown in \cref{fig:gammas} right.
Some features are common to the constraints from dwarf galaxies, for instance those associated to the resonances of the annihilation cross-section. Light DM candidates are not excluded mainly for two reasons: the cross-section (at fixed value of $\mVD$) decreases at smaller masses and we are focusing on gamma-ray energies above 1.5 GeV. Including the formation of bound states in the analysis allow to rule out an additional region of the parameter space at $\MDM \sim 1.5-10$ TeV.

\subsection{Antiproton searches}
\label{sec:antiproton}

The antiproton component in cosmic rays has been recognized since decades as a potentially important channel for DM searches. The recent release~\cite{Aguilar:2016kjl} of high-precision data by the {\sc Ams-02} experiment concerning the antiproton flux and antiproton to proton ratio has on one side strengthened the relevance of antiprotons and on the other side pointed to a need for a better determination of the astrophysical background (called `secondary antiprotons'). At the moment, such background can explain the data, within the rather large uncertainties, without the need for an additional exotic component.~\footnote{For recent studies on the subject, in some cases advocating a different conclusion, see~\cite{Evoli:2015vaa,Kappl:2015bqa,Cuoco:2016eej,Cui:2016ppb,Feng:2016loc,Huang:2016tfo}.} Hence, it remains meaningful to derive constraints based on antiprotons with the best tools currently at disposal. 

\medskip

We follow closely the analysis strategy of~\cite{Giesen:2015ufa}, which in turn builds on~\cite{Boudaud:2014qra}. We just recall here the most important features. {\sc Ams-02} has measured the flux in an energy range that extends from 1 to 450 GeV. The $\bar p/p$ ratio, which we will use for consistency with~\cite{Giesen:2015ufa}, shows a very good agreement with the predicted astrophysical background up to about 50-100 GeV and then remains rather flat in energy. The data can be reasonably well fit with secondaries assuming a {\sc Max} galactic propagation scheme and adapting the normalization and the impact of solar modulation within the uncertainties (we refer to \cite{Giesen:2015ufa} for any further detail). For the DM contribution, we choose for definiteness an Einasto profile for the distribution in the MW galactic halo and we include all the relevant propagation phenomena as discussed in~\cite{Boudaud:2014qra}. By requiring that the DM contribution does not worsens the secondary-only fit by more than $\Delta \chi ^2 = 9$, we can derive conservative DM bounds. 

\medskip

\begin{figure}[t]
\begin{minipage}{0.48\textwidth}
\begin{center}
\includegraphics[width= \textwidth]{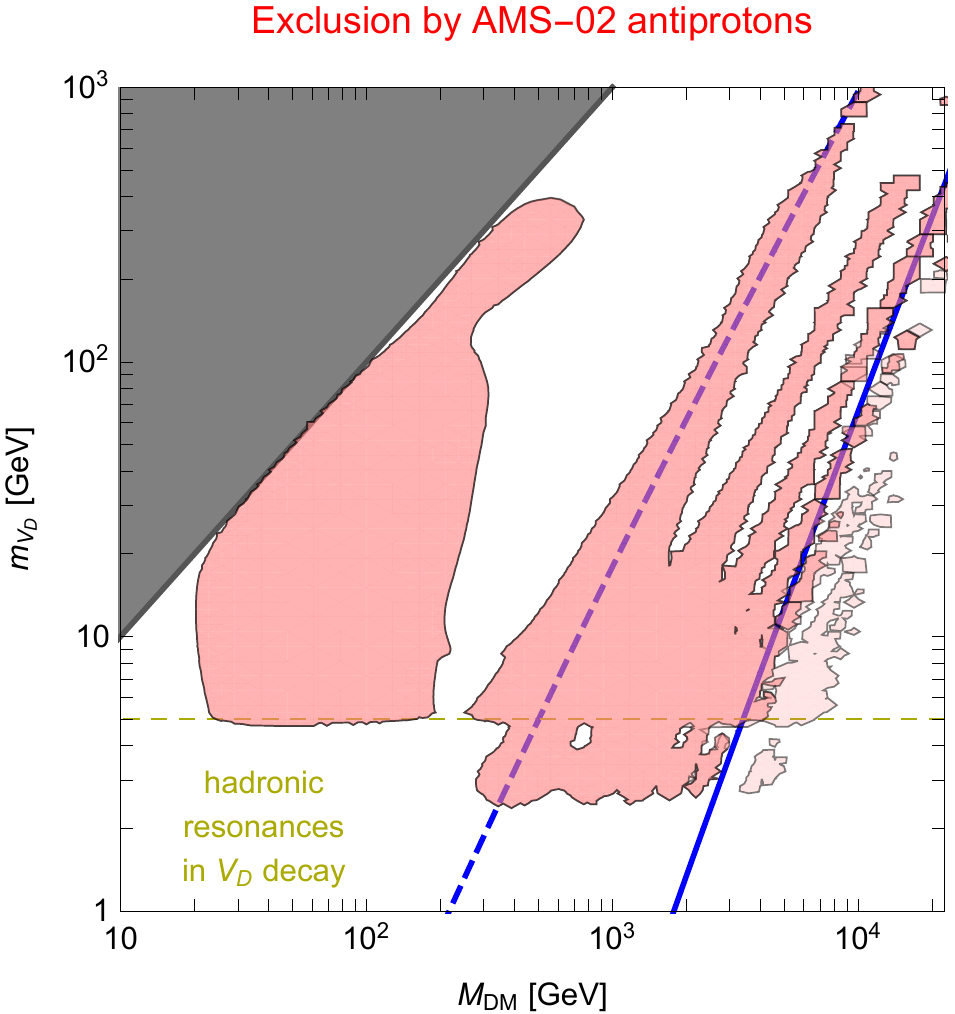}
\caption{\em \small \label{fig:pbar} {\bfseries Antiproton constraints} in the plane $\MDM/\mVD$, without Bound State Formation (darker shaded area) and with Bound State Formation (entire shaded area).}
\end{center}
\end{minipage}
\hspace{2mm}
\begin{minipage}{0.5\textwidth}
\begin{center}
\includegraphics[width= \textwidth]{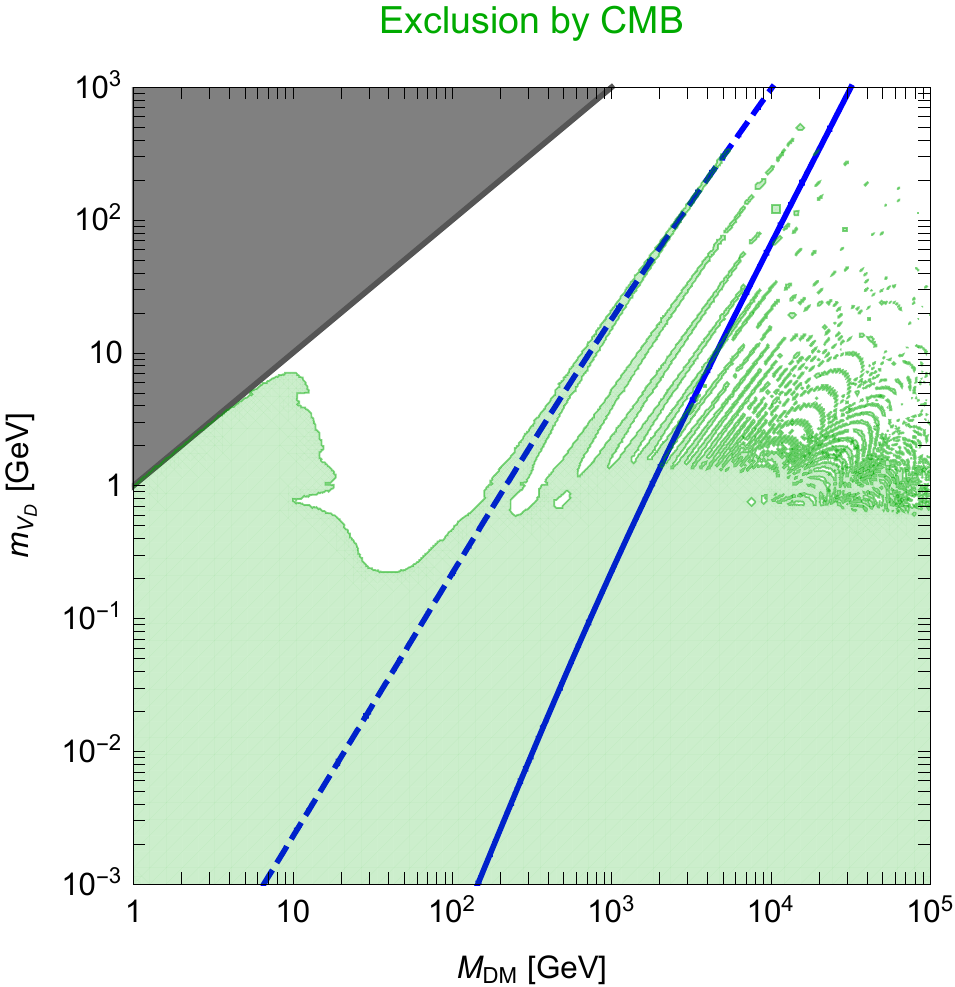}
\caption{\em \small \label{fig:CMB} {\bfseries CMB constraints}. At large DM masses the resonances are extremely dense. The small `crescent' features of the shading are just an artefact of the finite resolution of the plot.}
\end{center}
\end{minipage}
\end{figure}

The constraints that we obtain are reported in \cref{fig:pbar}. With respect to the other probes that we consider, they cover smaller areas, necessarily located above $\mVD \gtrsim 2m_p$ where antiproton production from $\VD$ decays is kinematically allowed. More precisely, the excluded region consists of two separate areas. The low mass one, for $\MDM \lesssim 200$ GeV, corresponds to the case in which the DM $\bar p$ flux falls in the high precision part of the {\sc Ams-02} data: even a small DM component is constrained by the small experimental error bars. The high mass one ($300 \ {\rm GeV} \lesssim \MDM \lesssim 3$ TeV) corresponds instead to the case in which the large enhancements of the resonance peaks in the annihilation cross-section imply a large DM flux and therefore an overshooting of the measurements. 

We notice that the constraints in the low mass area rule out certain models which have been put forward to explain the GC GeV excess, featuring a DM with a mass around 30-50 GeV and dark photons heavier than 1 GeV (see e.g.~\cite{Berlin:2014pya,Boehm:2014bia}). While we caution that this is a complex region, where significant astrophysical uncertainties come into play~\cite{Cirelli:2014lwa,Giesen:2015ufa} and that would therefore require a dedicated study, we find it worthwhile to stress for the first time that the {\sc Ams-02} antiproton data have the power to test this portion. Naturally, models for the GC GeV excess where $\mVD \lesssim 1$ GeV remain unconstrained.

\medskip

The impact of adding the formation of bound states is limited for antiprotons: only a small additional region around $\MDM \simeq 7$ TeV, $\mVD \simeq 7$ GeV is excluded. This is not surprising: antiproton constraints necessarily appear only for $\mVD \gtrsim 2m_p$ and bound state formation only occurs to the right of the solid blue line, thus restricting the applicable area only to the upper-right portion of the parameter space. On the other hand, for DM masses larger than $\sim$10 TeV the constraints quickly lose power because the maximal energy of the {\sc Ams-02} data and because the annihilation cross section decreases.

\subsection{CMB constraints}
\label{sec:CMB}

\paragraph{Bound on the annihilation cross section.} 
We base our DM constraints on the recent detailed analysis of ref.~\cite{Slatyer:2015jla} (itself based on the inputs of~\cite{Slatyer:2015kla}), and on the {\sc Planck} measurements of CMB anisotropies~\cite{Ade:2015xua}.
We take $f_{\rm eff}$ from~\cite{Slatyer:2015jla}, and use it in conjunction with the bound of ref.~\cite{Ade:2015xua}, to derive constraints on our parameter space. We shall require that
\begin{equation}
\sum_{{\rm SM} = \bar{e}e, \bar{\mu}\mu, \dots}\langle \sigma_{\rm tot} v_{\rm rel}\rangle 
\ {\rm BR}_{\VD \to {\rm SM}} 
\ f_{\rm eff}(\MDM, {\rm SM}) 
< 8.2 \cdot 10^{-28} \, \frac{\rm cm^3}{\rm s} \left(\frac{\MDM}{\rm GeV} \right)\,,
\label{eq:CMBbound}
\end{equation}
where we remark that CMB bounds are insensitive to the number of steps of the cascade~\cite{Elor:2015bho}, in contrast to the indirect detection bounds of sections  \ref{sec:gammas} and \ref{sec:antiproton}.

In eq.~(\ref{eq:CMBbound}) $\langle \sigma_{\rm tot} v_{\rm rel} \rangle$ accounts for all processes via which DM annihilates. It depends, in general, on the DM speed distribution, and should be evaluated at the redshift that affects the ionisation of the medium most strongly. The effect on the ionisation has been shown to peak at $z \sim 600$ and extend down to $z \sim$~few~hundred~\cite{Finkbeiner:2011dx}.
We need therefore to determine the DM speed distribution at those redshifts.

\paragraph{Dark matter velocity around CMB.}
The DM temperature at redshifts relevant to CMB can be found from \cref{eq:TX,eq:ztrans,eq:TDtrans}. It indicates a Maxwellian DM speed distribution with central value $v_0 = \sqrt{2\TX/\MDM}$, where
\beq
v_0 \approx  10^{-8}
\ \left(\frac{1+z}{600}\right)
\sqrt{ \min\left[
\left(\frac{{\rm MeV}}{\mVD}\right) \left(\frac{\rm GeV}{\MDM}\right) , \
\left(\frac{\alphaD}{10^{-6}}\right) \left(\frac{\rm GeV}{\MDM}\right)^{5/2}
\right] } \,.
\label{eq:vCMB}
\eeq
Note that we have assumed a sharp transition between the  $T_X \propto 1/a$ and $T_X \propto 1/a^2$ scalings, which is more than sufficient for our purposes.
As always, for the relative velocity of the DM particles, we use $v_{{\rm rel}, 0} =\sqrt{2} \, v_0$.

\paragraph{Annihilation cross section.} 
We now turn to the evaluation of $\langle \sigma_{\rm tot} v_{\rm rel} \rangle$ entering the constraint \eqref{eq:CMBbound}.
\begin{itemize}
\renewcommand\labelitemi{$\diamond$}

\item
For the range of DM velocities indicated by \cref{eq:vCMB} and the range of DM and dark photon masses we consider in this work, only the direct DM annihilation contributes, while BSF is extremely suppressed due to its $\sigma_{\rm BSF} v_{\rm rel} \propto v_{\rm rel}^2$ scaling at low velocities, discussed in \cref{sec:cross-sections}. Hence, $\langle \sigma_{\rm tot} v_{\rm rel} \rangle \simeq \langle \sigma_{\rm ann} v_{\rm rel} \rangle$.

\item
Because $v_0$ is very low, the amplitude of the $S_{\rm ann}$ resonances is very large, and their numerical calculation is impractical. We use instead a standard analytical approximation obtained by replacing the Yukawa with the Hulth\'en potential,\footnote{
For the other indirect detection analysis in this section, we instead calculate $S_{\rm ann}$ 
fully numerically, using the Yukawa potential.} 
$V_H = -\alphaD m_* \, e^{-m_* r}/(1-e^{-m_* r})$, where $m_* \sim {\cal O}(\mVD)$. Then, the Sommerfeld enhancement factor is (see e.g.~\cite{Cassel:2009wt, Slatyer:2009vg})
\beq
S_H = \frac{2\pi \alphaD}{v_{\rm rel}} \,
\frac{\sinh(\pi \MDM v_{\rm rel}/m_*)}
{
\cosh(\pi \MDM v_{\rm rel}/m_*) - 
\cosh \left(\pi \sqrt{\MDM^2 v_{\rm rel}^2/m_*^2 - 4 \MDM \alphaD/m_*} \right)
} \,.
\label{eq:S_Hulthen}
\eeq
While $S_H$ is a good approximation off-resonance, it does not reproduce the precise position of the resonances of the Yukawa potential~\cite{Cassel:2009wt}, beyond the first one, whose position we can define. Indeed, we choose $m_* = 1.68 \, \mVD$, such that the first resonances of the Yukawa and the Hulth\'en potentials coincide.\footnote{This is slightly different from the more common choice found in the literature, $m_*/\mVD = \pi^2/6 \simeq 1.64$~\cite{Cassel:2009wt}.}
Despite the discrepancy in the location of the higher resonances, the Hulth\'en potential reproduces correctly the fact that they become denser at increasing values of $\alphaD\MDM/\mVD$, and is overall sufficient for our purposes.

It is well-known that at very low velocities, the growth of the resonances of the Hulth\'en (or the Yukawa) potential is unphysical, a fact that is manifested by the apparent violation of the unitarity limit on the inelastic cross-section. We use the prescription of ref.~\cite{Blum:2016nrz} to ensure that the resonant growth is curtailed below the unitarity limit. For this purpose, we replace the Sommerfeld enhancement factor $S_H$ with $S_H^{\rm reg}$, where\footnote{
Setting $w \equiv (\sigma v_{\rm rel})/(\sigma_{\rm uni} v_{\rm rel})$ and 
$w^{\rm reg} \equiv (\sigma^{\rm reg} v_{\rm rel})/(\sigma_{\rm uni} v_{\rm rel})$, 
where $\sigma^{\rm reg}$ is the regulated cross-section and $\sigma_{\rm uni}$ is the unitarity limit, this prescription can be re-expressed more generally as 
$w^{\rm reg} = w/(1+w/4)^2$, which shows clearly that $w^{\rm reg} < 1$ for any $w$.  
Note that the prescription is valid for $s$-wave annihilation only.
}
\beq
S_H^{\rm reg} = \dfrac{S_H}{\left(1 + S_H \, \alphaD^2 v_{\rm rel} / 16 \right)^2} \,.
\label{eq:Sreg}
\eeq

\item
In the parameter space of interest, $\sigma_{\rm ann} v_{\rm rel}$ is well within the saturated regime, \textit{i.e.} it is independent of $v_{\rm rel}$. It is then consistent to use the constraint \eqref{eq:CMBbound}, which applies to a velocity-independent $\sigma v_{\rm rel}$. Moreover, the precise value of the speed around $z \sim 600$ is unimportant, and there is no need to average over the DM speed distribution.

\end{itemize}

\paragraph{Results.} The portion of our parameter space excluded by CMB is shown as a shaded green area in \cref{fig:CMB}. Values of $\mVD \lesssim 1$~GeV are ruled out for any DM mass, with the exception of a `gap' at $\MDM \sim (10-200)$ GeV, that extends down to $m_\VD \sim 200$~MeV. 
For larger values of $\mVD$, and for $\MDM \gtrsim 1$~TeV, the excluded regions follow the resonances of the annihilation cross-section.
Our CMB bounds are in good agreement with those derived in ref.~\cite{Bringmann:2016din} for the same model, and extend to larger values of $\mVD$ and $\MDM$. Moreover, they broadly agree with those derived in~\cite{Slatyer:2015jla,Elor:2015bho}. We complement them by considering the dark photon mass as an explicit parameter and combining the different annihilation channels as dictated by the actual branching ratios of $\VD$.
For $\mVD \gtrsim 1$ GeV and $\MDM \gtrsim 1$ TeV, our inclusion of BSF in the DM relic density computation is important, because it alters the predicted $\alphaD$ with respect to a calculation that includes annihilation only (see \cref{sec:relic}). This, in turn, affects the position of the resonances (which is, however, difficult to visualize because they are very dense), but mildly so on the overall size of the excluded area.

\section{Conclusions}
\label{sec:conclusions}

\begin{figure}[!t]
\begin{center}
\includegraphics[width= 0.67 \textwidth]{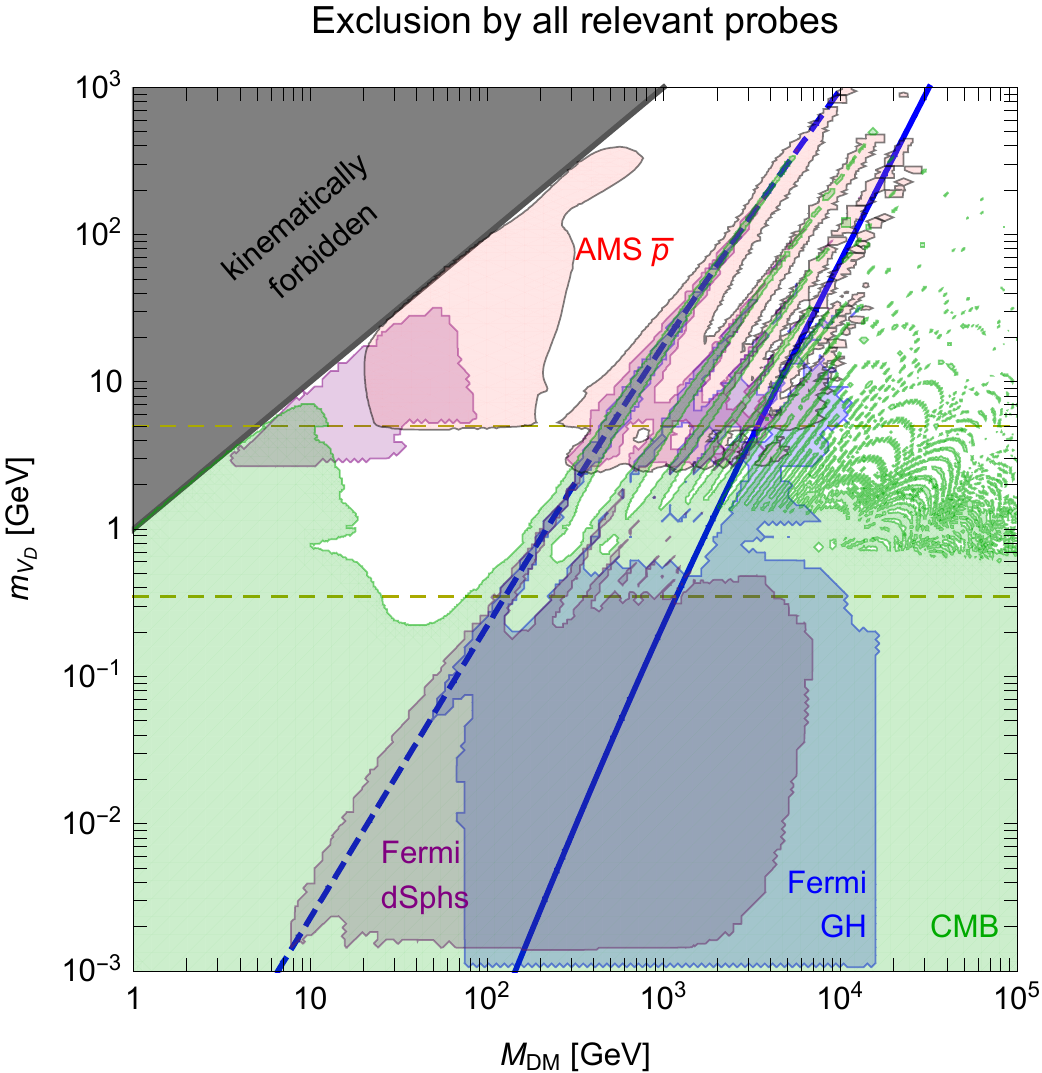}
\caption{\em \small \label{fig:allbounds} {\bfseries Compilation of all ID constraints} in the plane $\MDM/\mVD$, when including Bound State Formation.}
\end{center}
\end{figure}

Dark $U(1)$ sectors are ubiquitous in beyond-the-Standard-Model physics, ranging from string-theory constructions to supersymmetric extensions, and models with phenomenological motivation. Such sectors often involve stable particles that may have been produced with cosmologically significant abundances in the early universe, and make up part or all of the DM today. In this work, we considered DM as a Dirac fermion coupled to a dark $U(1)_D$ force, and carried out a comprehensive study of its phenomenology.

As with all relics from the early universe, the phenomenology of DM today depends on the preceding cosmology. For the first time in the  literature, the present work incorporates self-consistently, in the cosmology and the phenomenology of this model, effects that had been previously only separately considered. In particular, we accounted for the effect of bound states on the DM relic density --thus on the estimated dark sector couplings-- and computed direct and indirect detection constraints on the model parameters. 

Our main points can be summarized as follows:
\begin{itemize}
\renewcommand\labelitemi{$\star$}

\item
We identify the regions of the parameter space of the model in which bound state formation is relevant (\cref{sec:cross-sections}) and quantify precisely its impact. In these regions, BSF increases the total effective annihilation rate and therefore allows to obtain the observed DM density with a lower dark coupling. Neglecting the effect of BSF on the relic density can result in an error 
up to a factor of $\sim 4$ in the expected signal strength, for a given DM mass.

\item
Despite the decrease in the predicted coupling, the expected indirect detection signals are increased, due to the BSF contribution, leading to the exclusion of additional regions of the parameter space (see \cref{fig:gammas,fig:pbar}).

\item
We detailed the cosmological history of the dark sector, and estimated the constraints implied from not disrupting BBN and changing the time of matter-radiation equality (\cref{sec:BBN}). These constraints complement the direct and indirect detection ones, as they put a lower bound on the coupling between the dark sector and the SM, and imply that this coupling may be vanishingly small only for very light (eV-scale) or massless dark photons, whose density is redshifted away by the expansion of the universe. Bounds from extra radiation during CMB are instead less constraining~\cite{Ade:2015xua} (see ref.~\cite{Vagnozzi:2017ovm} for related discussion).

\item

MeV -- GeV dark-photon masses, which are relevant for the leptophilic realizations of the model, are highly constrained by the combined cosmological, direct and indirect detection bounds for any DM mass in our range (see \cref{fig:allbounds,fig:eps_mVD}). This sub-GeV dark-photon mass range has also been invoked in the context of self-interacting dark matter. The bounds could be evaded if the dark photons are very light or massless, $\mVD \lesssim 15$~eV (see e.g.~\cite{Feng:2009mn,Agrawal:2016quu}), and/or DM possesses a particle-antiparticle asymmetry~\cite{CyrRacine:2012fz,Cline:2013pca,Petraki:2014uza,Boddy:2016bbu}, or perhaps if the dark plasma never thermalised, i.e.~no significant cosmological abundance of dark photons was produced, while DM was populated due to pair-production processes by the SM fermions via the kinetic mixing of the dark photon to hypercharge (freeze-in)~\cite{Chu:2011be,Bernal:2015ova}.

\item
Dark photons heavier than 1 GeV are somewhat less constrained in the context of the present model, with the antiproton measurements and the CMB being the most powerful probes in this regime. The antiproton bounds put under strain models that explain the GC GeV excess in terms of 40 GeV DM coupled to heavy (above GeV) dark photons (although astrophysical uncertainties make a definite statement challenging).

\item
In the multi-GeV dark photon region, the indirect detection constraints of \cref{fig:allbounds} would be weakened for a very small value of the kinetic mixing $\epsilon$. Small $\epsilon$ values (between $10^{-9}$ and $10^{-12}$, depending on the $m_\VD$ considered), while still guaranteeing a dark photon lifetime shorter than $\tauBBN$, would cause $\VD$ to decay when it dominates the energy density of the universe, thus injecting significant entropy in the SM plasma and diluting DM. The dilution factors are shown in \cref{fig:cosmology_bounds}. The dilution of DM would imply a smaller coupling $\alphaD$, in order to account for the observed DM density, and in turn the possibility to have DM of thermal origin heavier than $\sim 100$~TeV without violating the unitarity bound. We leave the study of the related phenomenology for future work.

Moreover, if in the early universe the dark plasma was at a significantly lower temperature than assumed in this work, $\alphaD$ would be expected to be smaller, in the entire parameter space considered. This would relax the CMB, direct and indirect detection constraints, but would also shift the parameter space within which the DM self-interactions are sizeable.

\end{itemize}

\appendix

\section*{Appendix}
\section{Coulomb approximation in relic density calculations \label{app:RelicDensity_Coulomb}} 

To ensure the validity of our computation of the DM relic density, we consider the time of \emph{chemical decoupling}, when the DM density has approached its final value, $Y_X \simeq \Omega_{\rm DM} \rho_c / (2 \MDM s_0)$, where $\rho_c$ and $s_0$ are the critical energy density and the entropy density today. Because of the Sommerfeld effect, the annihilation and BSF processes remain sizeable after freeze-out. While freeze-out -- conventionally defined as the time when the density of the DM particles departs from its equilibrium value -- occurs at $\TD^{\rm f.o.} \approx \MDM / 30$, the chemical decoupling may occur significantly later. 

We shall define the chemical decoupling as the time when the rate of change of the DM comoving density becomes $|d\ln Y_X / d \xD| \lesssim 1\%$, where $\xD\equiv \MDM/\TD$. Around that time, the DM density is governed by the Boltzmann equation
$d\ln Y_X / d\xD \approx - \sqrt{g_{*}} 
[\sigma_0/(1.7 \times 10^{-26}{\rm cm^3/s})] 
\, [S_{\rm eff} (\xD)/\xD^2]$, 
where on the right side of the equation, we have substituted the final value of the DM density. The ``effective" Sommerfeld factor $S_{\rm eff}$ is defined as in ref.~\cite{vonHarling:2014kha}; it includes the thermally averaged Sommerfeld enhancement factor $\langle S_{\rm ann}\rangle$, as well as $\langle S_{\rm BSF}\rangle$ but only after the bound-state ionisation rate has become lower than the bound-state decay rate. For the precise definition of $\sqrt{g_*}$, see ref.~\cite{Baldes:2017gzw}. Using these equations, we determine the temperature of chemical decoupling $\TD^{\rm c.d.}$, for a given $\MDM$ and the associated $\alphaD$.

The condition \eqref{eq:CoulombCondition} for the Coulomb approximation is fulfilled provided that the chemical decoupling occurs at $\TD^{\rm c.d.} \gtrsim \mVD^2 / \MDM$. We find that this is satisfied for the $\mVD$ range below the upper dotted red line in \cref{fig:PhaseSpace}. Since this encompasses essentially all of the parameter where the Sommerfeld effect is important (below dashed blue line), including the entire region where bound states can form, we deem the Coulomb approximation to be satisfactory for the computation of the relic density, in the entire region where $S_{\rm ann}, S_{\rm BSF} > 1$.

Our computation also introduces a Coulombic Sommerfeld enhancement in the parameter space above the dashed blue line of \cref{fig:PhaseSpace}, where in fact $S_{\rm ann} \simeq 1$ at any speed, due to $\mVD$ being large. This raises the concern that $\sigma_{\rm ann} v_{\rm rel}$ may be overestimated in this regime. However, for $\MDM \lesssim$~TeV, which encompasses most of the parameter space above the dashed blue line, the Sommerfeld enhancement has no significant effect on the DM density, even when its Coulomb limit is employed, because $\alphaD/v_{\rm rel}$ remains small before chemical decoupling. Thus, including a Coulombic Sommerfeld enhancement in the computation does not affect the predicted $\alphaD$ in this region, and our results remain valid above the dashed blue line of \cref{fig:PhaseSpace} as well.

\section*{Acknowledgements}
We thank Iason Baldes, Francesc Ferrer, Andreas Goudelis, Felix Kahlhoefer, Bradley Kavanagh, Sunny Vagnozzi, Tien-Tien Yu and Gabrijela Zaharijas for interesting discussions. We acknowledge the hospitality of the Institut d'Astrophysique de Paris ({\sc Iap}) where a part of this work was done.

\medskip

{\small

\noindent Funding and research infrastructure acknowledgements: 
\begin{itemize}
\renewcommand\labelitemi{$\ast$}
\setlength\itemsep{0ex}
\item European Research Council ({\sc Erc}) under the EU Seventh Framework Programme (FP7/2007-2013)/{\sc Erc} Starting Grant (agreement n.\ 278234 --- `{\sc NewDark}' project).
\item  French state funds managed by the {\sc Anr} (Agence Nationale de la Recherche), in the context of the {\sc LabEx} ILP (ANR-11-IDEX-0004-02, ANR-10-LABX-63).
\item {\sc Anr} (Agence Nationale de la Recherche) under contract ACHN 2015 `{\sc The Intricate Dark}' project [work of K.P.].
\item {\sc Nwo} (The Netherlands Organisation for Scientific Research), VIDI grant, `{\sc Self-Interacting Asymmetric Dark Matter}' project [work of K.P.].
\item {\sc Erc} Advanced Grant project 267117 (`{\sc Dark}') hosted by Universit\'e Pierre et Marie Curie -- Paris 6 [initial work of P.P.].
\item Centro de Excelencia Severo Ochoa Programme SEV-2012-0249 [work of M.T.].
\item MINECO through a Severo Ochoa fellowship with the Program SEV-2012-0249 [work of M.T.].
\item FPA2015-65929-P and Consolider MultiDark CSD2009-00064 [work of M.T.].
\end{itemize}

}

\def\bibfont{\footnotesize}\setlength{\bibsep}{3pt}
\providecommand{\href}[2]{#2}\begingroup\raggedright\endgroup

\end{document}